\documentclass{aa}
\usepackage{graphicx}
\usepackage[varg]{txfonts}
\usepackage{color}
\usepackage{longtable}
\usepackage{amsmath}
\usepackage{natbib,twoopt}
\usepackage[breaklinks=true]{hyperref} %% to avoid \citeads line fills
\bibpunct{(}{)}{;}{a}{}{,}             %% natbib format for A&A and ApJ
\makeatletter
  \newcommandtwoopt{\citeads}[3][][]{\href{http://adsabs.harvard.edu/abs/#3}%
    {\def\hyper@linkstart##1##2{}%
     \let\hyper@linkend\@empty\citealp[#1][#2]{#3}}}
  \newcommandtwoopt{\citepads}[3][][]{%
  \nonstopmode%
   \href{http://adsabs.harvard.edu/abs/#3}%
    {\def\hyper@linkstart##1##2{}%
     \let\hyper@linkend\@empty\citep[#1][#2]{#3}}%
   \errorstopmode}
 \newcommandtwoopt{\citetads}[3][][]{%
 \nonstopmode%
  \href{http://adsabs.harvard.edu/abs/#3}%
  {\def\hyper@linkstart##1##2{}%
  \let\hyper@linkend\@empty\citet[#1][#2]{#3}}%
  \errorstopmode}
  \newcommandtwoopt{\citeyearads}[3][][]%
    {\href{http://adsabs.harvard.edu/abs/#3}
    {\def\hyper@linkstart##1##2{}%
     \let\hyper@linkend\@empty\citeyear[#1][#2]{#3}}}
\makeatother

\begin{document}

\title{Inhomogeneous molecular ring around the B[e] supergiant 
LHA\,120-S\,73\thanks{Based on observations (1) with the 1.52-m and 2.2-m 
    telescopes at the European Southern Observatory (La Silla, Chile), under 
    the programme 076.D-0609(A) and under the agreement with the 
    Observat\'orio Nacional-MCT (Brazil), (2) at the 
    Gemini Observatory, which is operated by the Association of Universities 
    for Research in Astronomy, Inc., under a cooperative agreement with the
    NSF on behalf of the Gemini partnership: the National Science
    Foundation (United States), the Science and Technology Facilities
    Council (United Kingdom), the National Research Council (Canada),
    CONICYT (Chile), the Australian Research Council (Australia),
    Minist\'{e}rio da Ci\^{e}ncia, Tecnologia e Inova\c{c}\~{a}o (Brazil)
    and Ministerio de Ciencia, Tecnolog\'{i}a e Innovaci\'{o}n Productiva
    (Argentina), under the programmes GS-2004B-Q-54, GS-2010B-Q-31, and
    GS-2012B-Q-90, (3) at Complejo
Astron\'{o}mico El Leoncito operated under agreement between the Consejo
Nacional de Investigaciones Cient\'{i}ficas y T\'{e}cnicas de la Rep\'{u}blica 
Argentina and the National Universities of La Plata, C\'{o}rdoba and San Juan 
(Visiting Astronomer: A.F.T.), and (4) with the du Pont 
Telescope at Las Campanas Observatory, Chile, under the programme CNTAC 2008-02
(Visiting Astronomer: R.B. and A.F.T.).
} \thanks{Presented spectroscopic data are available in electronic form
at the CDS via anonymous ftp to cdsarc.u-strasbg.fr (130.79.128.5)
or via http://cdsweb.u-strasbg.fr/cgi-bin/qcat?J/A+A/}}

\author{M.~Kraus\inst{1,2}, L.~S.~Cidale\inst{3,4}, M.~L.~Arias\inst{3,4}, 
G.~Maravelias\inst{1}, D.~H.~Nickeler\inst{1}, A.~F.~Torres\inst{3,4}, 
M.~Borges Fernandes\inst{5}, 
A.~Aret\inst{2}, 
M.~Cur\'{e}\inst{6}, R.~Vallverd\'u\inst{3,4}, R.~H.~Barb\'{a}\inst{7}}

\institute{
Astronomick\'y \'ustav, Akademie v\v{e}d \v{C}esk\'e republiky,
Fri\v{c}ova 298, 25165 Ond\v{r}ejov, Czech Republic\\
\email{michaela.kraus@asu.cas.cz}
\and
Tartu Observatory, T\~oravere, 61602 Tartumaa, Estonia
\and
Departamento de Espectroscop\'ia Estelar, Facultad de Ciencias
Astron\'omicas y Geof\'isicas, Universidad Nacional de La Plata (UNLP), Paseo del Bosque s/n, B1900FWA, La Plata, Argentina
\and
Instituto de Astrof\'isica de La Plata, CCT La Plata, CONICET-UNLP,
Paseo del Bosque s/n, B1900FWA, La Plata, Argentina
\and
Observat\'orio Nacional, Rua General Jos\'e Cristino 77,
20921-400 S\~ao Cristov\~ao, Rio de Janeiro, Brazil
\and
Instituto de F\'{i}sica y Astronom\'{i}a, Facultad de Ciencias, Universidad
de Valpara\'{i}so, Av. Gran Breta\~na 1111,
Casilla 5030, Valpara\'{i}so, Chile
\and
Departamento de F\'{i}sica y Astronom\'{i}a, Universidad
de La Serena, Cisternas 1200 Norte, La Serena, Chile
}

\date{Received; accepted}

\authorrunning{Kraus et al.}
\titlerunning{Inhomogeneous molecular ring around LHA\,120-S\,73}

\abstract 
{B[e] supergiants are evolved massive stars, enshrouded in a dense 
wind and surrounded by a molecular and dusty disk. The mechanisms that drive 
phases of enhanced mass loss and mass ejections, responsible for the shaping of 
the circumstellar material of these objects, are still unclear.}
{We aim to improve our knowledge on the structure and dynamics of the 
circumstellar disk of the Large Magellanic Cloud B[e] supergiant 
LHA\,120-S\,73.}
{High-resolution optical and near-infrared spectroscopic data were 
obtained over a period of 16 and 7 years, respectively. The spectra cover the 
diagnostic emission lines from [\ion{Ca}{ii}] and [\ion{O}{i}], as well as the 
CO bands. These features trace the disk at different distances from the star. 
We analyzed the kinematics of the individual emission regions by modeling their 
emission profiles. A low-resolution mid-infrared spectrum was obtained as well, 
which provides information on the composition of the dusty disk.}
{All diagnostic emission features display double-peaked line profiles, which we 
interpret as due to Keplerian rotation. We find that the profile of each
forbidden line contains contributions from two spatially clearly distinct 
rings. In total, we find that LHA\,120-S\,73 is surrounded by at least four 
individual rings of material with alternating densities (or by a disk with 
strongly non-monotonic radial density distribution). Moreover, we find that 
the molecular ring must have gaps or at least strong density inhomogeneities, 
or in other words, a clumpy structure. The optical spectra additionally display a broad 
emission feature at 6160--6180\,\AA, which we interpret as molecular emission
from TiO. The mid-infrared spectrum displays features of oxygen- and
carbon-rich grain species, which indicates a long-lived, stable dusty disk.
 
We cannot confirm the previously reported high value for the stellar rotation velocity.
\ion{He}{i} $\lambda$ 5876 is the only clearly detectable pure atmospheric 
absorption line in our data. Its line profile is strongly variable in both 
width and shape and resembles of those seen in non-radially pulsating stars. A 
proper determination of the real underlying stellar rotation velocity is hence
not possible.}
{The existence of multiple stable and clumpy rings of alternating density  
recalls ring structures around planets. Although there is
currently insufficient observational evidence, it is tempting to propose a 
scenario with one (or more) minor bodies or planets revolving
around LHA\,120-S\,73 
and stabilizing the ring system, in analogy to the shepherd moons in planetary
systems.}


\maketitle

\section{Introduction}
\label{sec:intro}

B[e] supergiants (B[e]SGs) are evolved massive stars 
\citepads{1986A&A...163..119Z}. They form a subgroup of objects displaying the 
B[e] phenomenon \citepads{Lamers1998}. Their optical spectra display a hybrid 
character, meaning that they present broad and intense Balmer emission lines and 
simultaneously narrow emission lines of low-excitation permitted and forbidden 
transitions of low-ionized and neutral elements (i.e., \ion{Fe}{ii}, 
[\ion{Fe}{ii}] and [\ion{O}{i}]). In addition, the stars exhibit a very strong 
near- or mid-infrared excess.

%%%%%%%%%% Tab. 1 %%%%%%%%%%%%%%%%%%%%%%%%

\begin{table*}
\caption{Observation summary.}
\label{tab:obs}
\centering
\begin{tabular}{lccccccr}
\hline 
\hline 
Date & $t_{\rm exp}$ & Instrument/ & Resolution & Wavelength \\ 
(UT) & (s)           & Observatory &            & Range      \\ 
\hline
1999-12-17 & 4500 & FEROS/ESO-La Silla & 55 000 & $3600-9200\,\AA$ \\ 
2005-12-12 & 1500 & FEROS/ESO-La Silla & 55 000 & $3600-9200\,\AA$ \\ 
2014-11-28 & 1600 & FEROS/ESO-La Silla & 55 000 & $3600-9200\,\AA$ \\ 
2015-10-12 & 1400 & FEROS/ESO-La Silla & 55 000 & $3600-9200\,\AA$ \\ 
2008-11-13 & 1800 & Echelle-du Pont/LCO & 45 000 & $3480-10150\,\AA$ \\ 
2006-01-16 & 2700 & REOSC/CASLEO & 13 000 & $4560-7100\,\AA$ \\ 
2012-11-26 & 2700 & REOSC/CASLEO & 13 000 & $5780-9100\,\AA$ \\ 
2012-11-27 & 1800 & REOSC/CASLEO & 13 000 & $5780-9100\,\AA$ \\ 
2004-12-17 & 1800 & Phoenix/GEMINI-South & 50 000 & $2.290-2.299\,\mu$m \\ 
2010-12-24 & 5920 & Phoenix/GEMINI-South & 50 000 & $2.290-2.299\,\mu$m \\ 
2011-01-04 & 5920 & Phoenix/GEMINI-South & 50 000 & $2.319-2.330\,\mu$m \\ 
2012-10-03 &  400 & T-ReCS/GEMINI-South  & 100 & $8.0-13.0\,\mu$m \\
\hline
\end{tabular}
\end{table*}

The B[e] phenomenon is typically related to the physical conditions of
the gaseous and dusty shells, rings, or disks surrounding a hot star and not to the properties of the star itself.
The origin of circumstellar envelopes of B[e]SGs is attributed to the
mass lost from the star either through dense, aspherical stellar winds or through 
sudden mass ejections expected to occur during short-lived phases in the 
post-main-sequence evolution of the stars. The possible structure of the 
B[e]SG circumstellar
envelopes was described by \citetads{1985A&A...143..421Z}, who proposed an
empirical model, consisting of a hot and fast line-driven wind in the polar
regions, and a slow, much cooler and denser (by a factor of $10^2-10^3$) wind
in the equatorial region. 

In this cool and dense equatorial region, the material condenses into molecules
and dust. Molecular emission from carbon monoxide (CO), tracing the inner rim of
the molecular region, was found in many B[e]SGs \citepads{1988ApJ...324.1071M,
1988ApJ...334..639M, 1989A&A...223..237M, 1996ApJ...470..597M,
2010MNRAS.408L...6L, 2012A&A...548A..72C, 2012MNRAS.426L..56O,
2013A&A...558A..17O, 2012A&A...543A..77W, 2013A&A...549A..28K}. Emission
from less prominent molecules such as titanium oxide
\citepads[TiO,][]{1989A&A...220..206Z, 2012MNRAS.427L..80T} and silicon
oxide \citepads[SiO,][]{2015ApJ...800L..20K} was also reported for a few objects.
In the outer parts of these equatorial regions, where the temperature drops
below the dust condensation value, dust forms. The large amount of dust
connected with B[e]SGs is visible as strong near- and mid-infrared excess
emission \citepads{1986A&A...163..119Z, 2009AJ....138.1003B,
2010AJ....140..416B} and resolved spectral dust features 
\citepads{2006ApJ...638L..29K, Kastner2010}. Interferometric observations of 
some Galactic B[e]SGs revealed that the dust is concentrated in a ring 
\citepads{2007A&A...464...81D, 2011A&A...525A..22D, 2012A&A...548A..72C}. In a 
few cases, the central object turned out to be a close binary, and the dusty
rings are circumbinary \citepads{2011A&A...526A.107M, 2012A&A...538A...6W, 
2012A&A...545L..10W}. 

Investigations of the kinematics within the gaseous (atomic and molecular)
disk regions often revealed that it is consistent with Keplerian rotation
\citepads{2011A&A...526A.107M, 2012MNRAS.423..284A, 2012A&A...548A..72C, 2012A&A...543A..77W, 2010A&A...517A..30K, 2013A&A...549A..28K, 2014ApJ...780L..10K, 
2015ApJ...800L..20K, 2015AJ....149...13M}. In some cases, observations
support evidence of disk variability as seen in LHA\,115-S\,18 
\citepads{2012MNRAS.427L..80T} and HD\,327083 \citepads{2013msao.confE.160K}, 
sudden disk formation as in LHA\,115-S\,65 \citepads{2012MNRAS.426L..56O}, and disk dissipation as in CI\,Cam \citepads{2014MNRAS.443..947L}.

To improve our knowledge on the disk formation process, the physical properties 
of the disks, and the timescales of possible variability and its 
origin, extensive observational monitoring of individual objects is 
indispensable. In the present work, we focus on the Large Magellanic Cloud
(LMC) B[e]SG star LHA\,120-S\,73. We investigate its gaseous (atomic and 
molecular) environment based on multi-epoch high-resolution optical and 
near-infrared spectroscopic observations and its dusty disk based on 
low-resolution mid-infrared data.

\section{Star}
\label{sec:star}
{\bf LHA\,120-S\,73} (HD\,268\,835, RMC\,66, IRAS\,04571-6954, 
HIP\,22\,989, $\alpha$ = $04^{\mathrm{h}}\,56^{\mathrm{m}}\,47\,\fs 077$ and
$\delta$ = $\,-69^{\degr}\,50^{\arcmin}$ $24\,\farcs 79$; J2000) is a bright 
emission-line star in the LMC \citepads{Henize1956}. It has previoiusly been 
classified by \citetads{Feast1960} as a peculiar A-type emission-line star.
\citetads{Fehrenbach1969} reported on numerous 
forbidden emission lines of singly ionized iron. Strong infrared excess emission
was found by \citetads{Stahl1983}. These authors also classified the object as a 
massive late B-type supergiant, surrounded by a dense envelope, which mimics an 
equivalent 
spectral type of an early A-type star, and a dusty shell. \citetads{Stahl1986} 
discovered narrow nebular emission lines, such as [\ion{O}{i}] and 
[\ion{N}{ii}] lines. As one of the dusty peculiar emission-line stars in the 
Magellanic Clouds, it was recognized as a B[e]SG by 
\citetads{1986A&A...163..119Z}, who introduced this designation.
 
The stellar parameters derived by \citetads{Stahl1983} are
$T_{\rm {eff}}$ = 12\,000\,K, $M_{\rm{bol}} = -8.94$\,mag, $L = 3\times 
10^{5}$\,L$_\odot$, $R$ = 125\,R$_\odot$, $M$ = 30\,M$_\odot$, and $E(B-V)$ = 
0.12. Similar values were derived by \citetads{vanGenderen1983}: 
$T_{\rm {eff}}$ = 12\,000\,K, $M_{\rm{bol}} = -9.4$\,mag, $L = 4.4\times 
10^{5}$\,L$_\odot$, $R$ = 143\,R$_\odot$, $M$ = 37\,M$_\odot$, and $E(B-V)$ = 
0.26. The wind terminal velocity was determined to $\varv_{\infty} \simeq
300$\,km\,s$^{-1}$ based on the observed low-resolution UV absorption
spectrum, and estimates for the mass-loss rate range from $\dot{M} = 6\times 
10^{-6}$\,M$_\odot$yr$^{-1}$ \citepads{1988A&A...190..103M} to $\dot{M} = 
3\times 10^{-5}$\,M$_\odot$yr$^{-1}$ \citepads{Stahl1983}.
   
%%%%%%%%%% Fig. 1 %%%%%%%%%%%%%%%%%%%%%%%%

\begin{figure*}[t]
\begin{center}
\includegraphics[width=\hsize,angle=0]{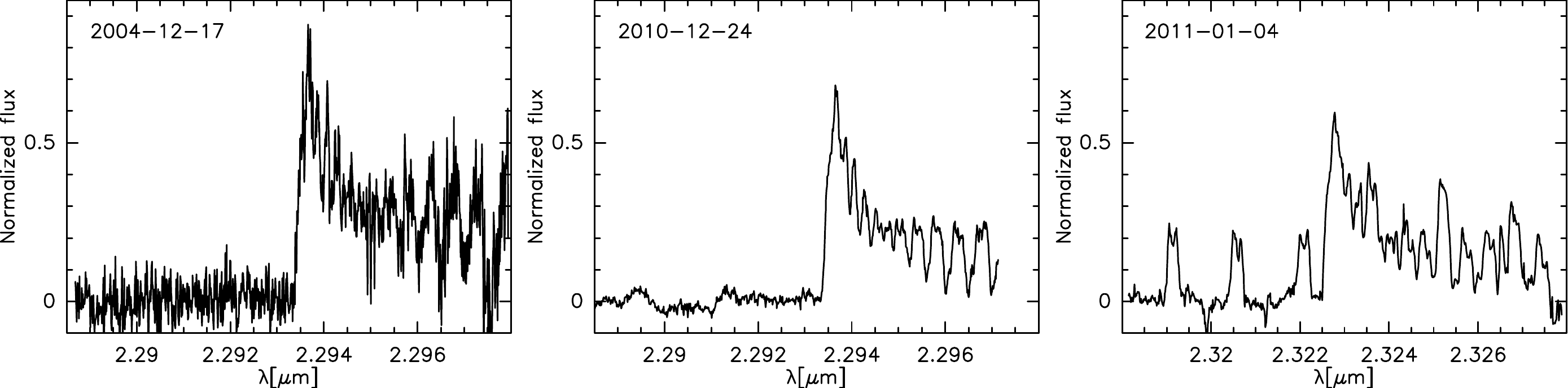}
\caption{Normalized and continuum-subtracted near-infrared Phoenix spectra of
LHA\,120-S\,73. Observations taken in 2004 (left) and 2010 (middle) cover
the first CO band head, those taken in 2011 (right) cover the second.}
\label{fig:s73_CO_obs}
\end{center}
\end{figure*}

Photometric observations by \citetads{2002A&A...386..926V} over a period of
about two years indicated slight variability. The authors identified two possible
periods, a long-term period ($\sim 380$ days) with an amplitude of
$\sim 0.06$\,mag superimposed on a short-term period ($\sim 55.5$ days) with an
amplitude of $\sim 0.03$\,mag. Based on the ASAS $V$-band light curve spanning 
ten years of observations, \citetads{2010AJ....140...14S} observed a 
pulsation-type oscillation with a period of $\sim 224$ days and an amplitude 
of $\sim 0.08$\,mag. This period is about four times longer than the 55.5 days
identified by 
\citetads{2002A&A...386..926V}. The periods were ascribed to possible
pulsations, and \citetads{2002A&A...386..926V} classified LHA\,120-S\,73 as
an $\alpha$\,Cyg variable. These authors mentioned that LHA\,120-S\,73
shares many similarities with luminous blue variables in an S\,Dor outburst
stage. Furthermore, \citetads{Stahl1983} and \citetads{Stahl1986} suggested that the
envelope might have been ejected during an S\,Dor-like outburst. However, no
S\,Dor typical variability was reported for LHA\,120-S\,73 so far.

The optical spectrum of LHA\,120-S\,73 exhibits countless narrow emission
lines, which display no shift with respect to the stellar radial velocity,
while the higher Balmer lines are blueshifted. The optical spectra were 
described in detail by \citetads{Stahl1983, 1985A&AS...61..237S}, who also 
found that the Balmer lines as well as some strong \ion{Fe}{ii} and the 
\ion{Ca}{ii}\,K lines display P\,Cygni profiles. Only a few typically very 
shallow photospheric absorption lines were reported.

Near- and mid-infrared excess emission is seen in the spectral energy 
distribution of 
LHA\,120-S\,73 \citepads{1986A&A...163..119Z, 2009AJ....138.1003B}, and 
observations from both {\it Spitzer} and {\it Herschel Space Telescopes} 
revealed that LHA\,120-S\,73 is one of nine evolved massive stars in the 
LMC with notable far-infrared emission \citepads{2010AJ....139...68V, 
2015ApJ...811..145J}. Moreover, emission features from amorphous and 
christalline silicates and from polycyclic aromatic hydrocarbons (PAHs) 
were detected in the data obtained with the {\it Spitzer Space Telescope} 
Infrared Spectrograph (IRS) \citepads{Kastner2010}. Intrinsic polarization due 
to dust was observed by \citetads{1992ApJ...398..286M},
suggesting a non-spherically symmetric distribution of the dust, probably 
confined to a disk. 

A disk seen close to pole-on was also inferred by 
\citetads{1986A&A...163..119Z} and \citetads{1988A&A...190..103M}, considering 
the huge amount of narrow permitted and forbidden emission lines of low-ionized 
metals (predominantly \ion{Fe}{ii}) coexisting with broad P\,Cygni absorption
components from a relatively high-velocity wind. \citetads{2012MNRAS.423..284A}
analyzed the kinematics of the two most valuable tracers of high-density 
gas: the lines of [\ion{O}{i}] and [\ion{Ca}{ii}]. They found that the
disk is rotating in Keplerian fashion and is seen under an inclination angle
of $28\pm\,1\degr$, in agreement with a more pole-on orientation of the star.

Near-infrared spectra of the star display pronounced, apparently stable 
molecular emission in both $^{12}$CO and $^{13}$CO 
\citepads{1988ApJ...334..639M, 2010MNRAS.408L...6L, 2013A&A...558A..17O}. The 
amount of identified $^{13}$CO implies that the surface of LHA\,120-S\,73 was 
strongly enriched in $^{13}$C at the time the material was released. The CO 
band emission typically originates from a (rotating) molecular gas ring, 
however, in these past observations the spectral resolution was too low to 
resolve the kinematics of the CO emitting region.

\section{Observations}
\label{sec:obs}

Mid- and near-infrared as well as optical spectra of LHA\,120-S\,73 were 
obtained. A summary of all observations is presented in Table\,\ref{tab:obs}, 
where we list the observing date (Col. 1), exposure time (Col. 2),
instrument and observatory (Col. 3), spectral resolution (Col. 4), and 
wavelength coverage (Col. 5).

\subsection{Mid-infrared spectrum}

A low-resolution ($R \sim 100$) mid-infrared spectrum in the $N$ band, covering
the range 8-13\,$\mu$m, was acquired in 2012 with the T-ReCS spectrograph
attached to the 8 m telescope at GEMINI-South (Chile). The spectrum was
processed using standard IRAF\footnote{IRAF is distributed by the National
Optical Astronomy Observatories, which are operated by the Association of
Universities for Research in Astronomy, Inc., under cooperative agreement with
the National Science Foundation.} tasks for mid-infrared spectroscopic data
reduction to obtain stacked images for the sky and difference frames, 
wavelength calibration, and to extract the target and
standard telluric star spectra.

Telluric correction of the target spectrum was performed using standard 
IRAF tasks. Finally, we obtained a rough flux calibration by 
multiplying the spectrum with a blackbody continuum of the same temperature as 
the telluric standard star. The flux scale was thereby determined from the 
available mid-infrared photometry of the science target.

%%%%%%%%%% Fig. 2 %%%%%%%%%%%%%%%%%%%%%%%%

\begin{figure*}[t]
\begin{center}
\includegraphics[width=\hsize,angle=0]{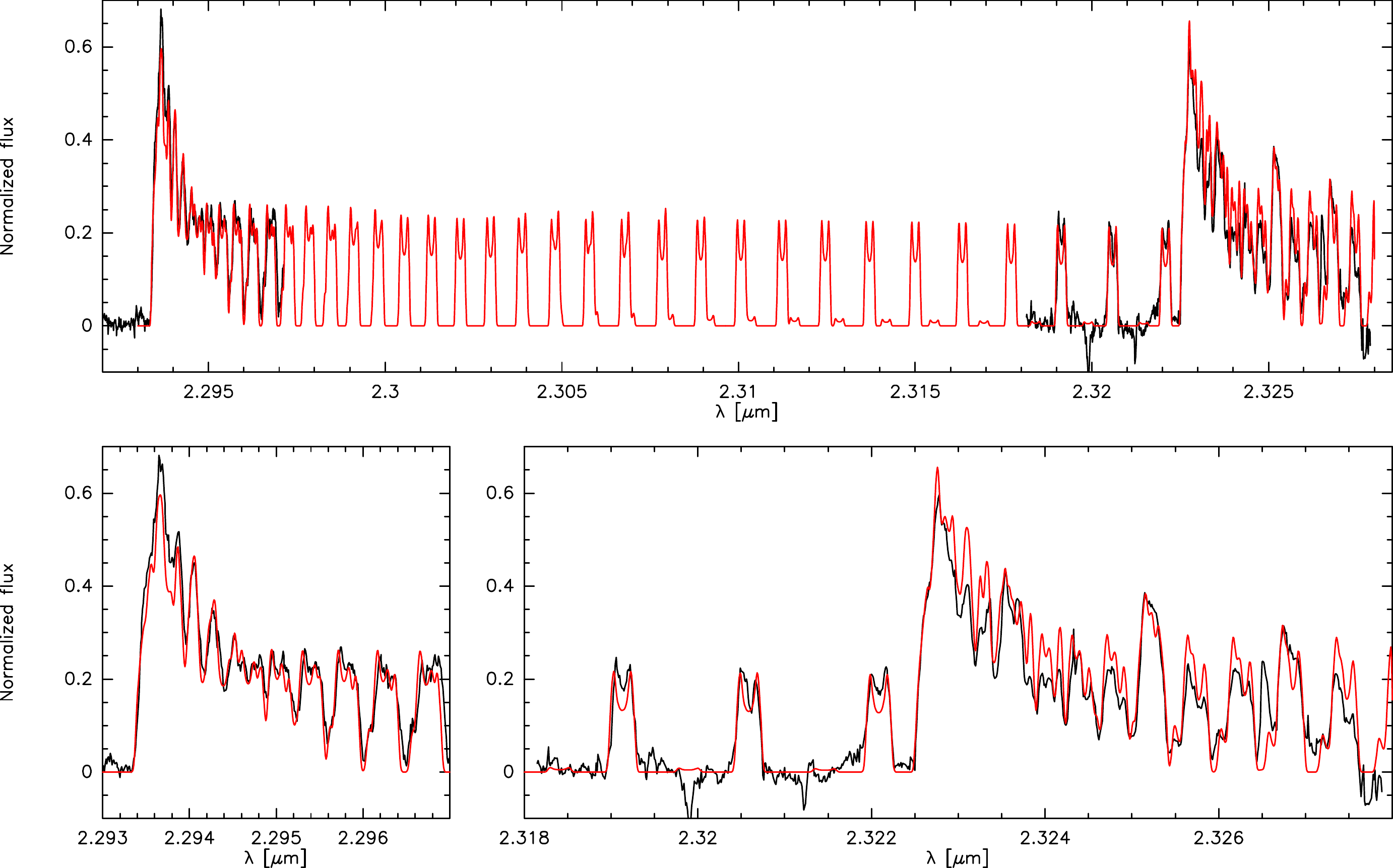}
\caption{Model fits (red) to the observed, normalized, and continuum-subtracted
(black) first and second CO band heads of LHA\,120-S\,73. The top panel shows
the complete CO band structure, the bottom panels only the observed band head
regions.}
\label{fig:s73-CO-fit}
\end{center}
\end{figure*}

\subsection{Near-infrared spectra}

We obtained high-resolution ($R\sim$50\,000) long-slit K-band spectra 
in December 2004, December 2010, and January 2011 with the 
visitor spectrograph Phoenix mounted at GEMINI-South (Chile). The spectra were 
obtained using two different filters, K4396 and K4308, covering a spectral 
range of $2.290-2.299\,\mu$m and $2.319-2.330\,\mu$m, respectively. This 
wavelength range was chosen because it covers the positions of the first and 
second band head of the rotation-vibrational transitions of the CO 
first-overtone bands.

A late-type B main-sequence telluric standard star was observed 
close in time to the target and at similar airmass. For ideal sky subtraction, 
the observations were taken in an ABBA nod pattern. Data reduction and 
telluric correction was performed using standard IRAF tasks.

The final, normalized, and continuum-subtracted near-infrared spectra are shown 
in Fig.\,\ref{fig:s73_CO_obs}. The data taken in 2004 are noisy, but the 
structure of the CO band head is still recognizable.

\subsection{Optical spectra}

Five medium-resolution spectra were obtained between 2006 and 2012 using the 
REOSC echelle spectrograph attached to the Jorge Sahade 2.15 m telescope at 
Complejo Astron\'{o}mico El Leoncito (CASLEO, Argentina). REOSC provides a 
resolving power of $R\sim$13000. The data from 2006 cover the spectral range 
4560-7100\,\AA \ and were taken with grating 580 with 400~l\,mm$^{-1}$. 
Those from 2012 cover the spectral range 5780-9100\,\AA \ and were observed 
with grating 180 with 316~l\,mm$^{-1}$. Data reduction was performed 
using standard IRAF tasks. In 2012, two spectra were taken each night, which 
were co-added for a better signal-to-noise ratio (S/N). 

In addition, four high-resolution spectra were obtained between 1999 and 2015
with the Fiber-fed Extended Range Optical Spectrograph
\citepads[FEROS][]{1999Msngr..95....8K}. During the
observations in December 1999, the spectrograph was attached to the
1.52 m telescope, while all other observations were carried out
with the spectrograph attached to the 2.2 m telescope, both at the European
Southern Observatory in La Silla (Chile). FEROS is a bench-mounted echelle
spectrograph with fibers, which cover a sky area of 2$\arcsec$ of diameter,
with a wavelength coverage from 3600\,\AA \ to 9200\,\AA \ and a spectral
resolution of R = 55\,000 (in the region around 6000 \AA). The data were
reduced using the FEROS reduction pipeline. A standard star was observed each
night for telluric correction, except for the night in 1999.
Telluric correction was performed using standard IRAF tasks, and for the data
from 1999, a telluric template from another night was applied.

Moreover, two high-resolution spectra were acquired in 2008 with the echelle
spectrograph attached to the 2.5 m du Pont telescope at Las Campanas
Observatory (LCO, Chile). The wavelength coverage ranges from 3480\,\AA \ to
10150\,\AA \ with a resolution of $\sim 45\,000$. A Tek5 2k$\times$2k CCD
detector with pixel size of 24 microns and a 1$\times$4$\arcsec$ slit were
used. The data were reduced using standard IRAF tasks. For these observations,
no telluric standard star was taken, hence the correction could not be
performed.

All spectra were corrected for the heliocentric velocity and for the
stellar radial velocity of 261\,km\,s$^{-1}$ determined by
\citetads{2012MNRAS.423..284A}.

\section{Results}
\label{sec:results}

\subsection{CO-band emission}
\label{sec:CO}

The high resolution of the near-infrared spectra unveils that
the CO band head structures display kinematical broadening. This is
obvious from the fully resolved double-peaked line profiles of the three
individual rotation-vibrational transitions blueward of the second band head
(right panel of Fig.\,\ref{fig:s73_CO_obs}). In addition, the first band head
displays a slightly broadened and asymmetric structure with a blue shoulder and
a red peak. This is characteristic of the emission originating in a
rotating molecular disk or ring \citepads{1993ApJ...411L..37C,
1995Ap&SS.224...25C, 1996ApJ...462..919N, 2000A&A...362..158K}. Hence, we
modeled the CO bands using the code developed by \citetads{2000A&A...362..158K}, which is suitable
to compute CO band emission in local thermodynamic equilibrium (LTE) from
a rotating disk. As was found in previous
investigations of CO bands in B[e]SGs, the molecular spectra are typically well
reproduced considering solely a narrow rotating ring because the observable CO
emission originates from the hot inner edge of the molecular disk
\citepads{2000A&A...362..158K, 2013A&A...549A..28K, 2009A&A...494..253K,
2010MNRAS.408L...6L, 2012A&A...548A..72C, 2015AJ....149...13M}. In addition,
the involved column densities in these rings are high, justifying the LTE
assumption. This
reduces the number of free model parameters to three. These are the temperature
of the CO ring $T_{\rm CO}$, the ring column density along the line of sight
$N_{\rm CO}$, and the deprojected rotation velocity of the molecular ring $\varv_{\rm rot}$.
For the inclination angle, we adopted the value of $28\degr$ estimated by
\citetads{2012MNRAS.423..284A}. To compare them with the observations, the
theoretical spectra were convolved with the resolution of the Phoenix
spectrograph.

%%%%%%%%%% Tab. 2 %%%%%%%%%%%%%%%%%%%%%%%%

\begin{table}
\caption{CO model parameters.}
\label{tab:CO-model}
\centering
\begin{tabular}{cccc}
\hline
\hline
$T_{\rm CO}$ & $N_{\rm CO}$ & $v_{\rm rot}$  & $i$ \\
   (K)       & (cm$^{-2}$)       & (km\,s$^{-1}$) & ($\degr$) \\
\hline
$2850\pm 100$ & $(6\pm 1)\times 10^{20}$ & $34\pm 1$ & 28 \\
\hline
\end{tabular}
\tablefoot{The inclination angle is taken from \citetads{2012MNRAS.423..284A} 
and is a fixed input value.}
\end{table}

%%%%%%%%%%%%%%%%%%%%%%%%%%%%%%%%%%%%%%%%%%%%%

%%%%%%%%%% Fig.3 %%%%%%%%%%%%%%%%%%%%%%%%

\begin{figure}[t]
\begin{center}
\includegraphics[width=\hsize,angle=0]{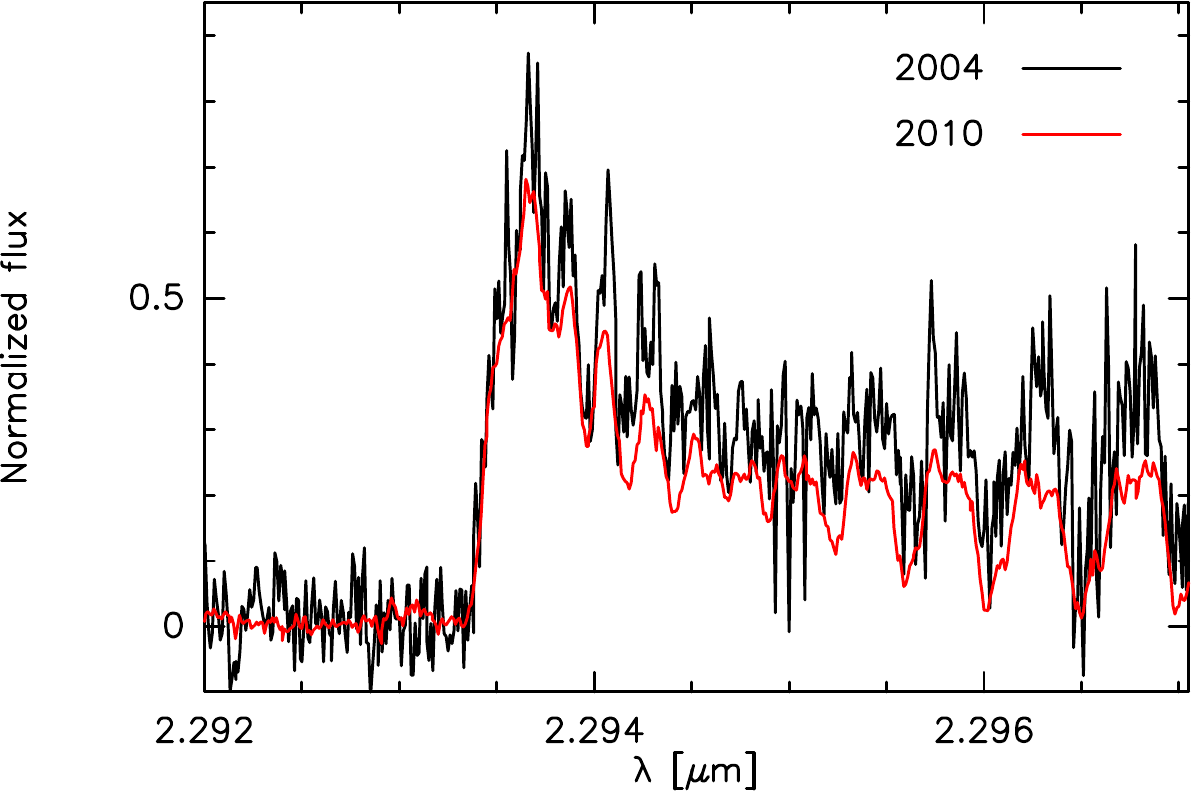}
\caption{Comparison of the intensities of the first CO band head observed
with Phoenix in 2004 and 2010.}
\label{fig:s73-variab}
\end{center}
\end{figure}

The quality of the 2004 spectrum of LHA\,120-S\,73 is poor. To determine the
physical parameters of the CO emitting region, we thus modeled the data taken in
2010/2011, which have the advantage of a much better quality, a longer
wavelength coverage (two band heads), and the spectra were taken close in time
with a separation of only 11 days.

Our spectra reveal no indication for emission from the hydrogen Pfund series.
This lack of Pfund lines is in agreement with former studies of
\citetads{2013A&A...558A..17O}, who found a contribution of the Pfund series in
LHA\,120-S\,73 only for wavelengths longer than 2.329\,$\mu$m. This red part of
the K-band spectrum is not covered by our observations. The minimum wavelength
of the Pfund line contribution found by \citetads{2013A&A...558A..17O}
corresponds to the line Pf(34), and this cutoff in the Pfund series at
relatively low quantum number implies a high electron density of $\sim
10^{15}$\,cm$^{-3}$ in the Pfund line emitting region \citepads[see,
e.g.,][]{2000Kraus}.

The best fit to the observed, normalized, and continuum-subtracted pure CO
band spectra is
shown in Fig.\,\ref{fig:s73-CO-fit}, and the fit parameters are listed in
Table\,\ref{tab:CO-model}. The upper panel of Fig.\,\ref{fig:s73-CO-fit}
shows the full computed spectral range in comparison to the observations,
while the lower panels highlight the two band head regions. Obviously,
there is some disagreement between the model and the observations. It is apparently impossible to fit
the strengths of both band head structures with a unique set of parameters.
Hence, the presented fit and its parameters provide a compromise, so that
the intensity of the first band head is not too strongly underestimated and
the one of the second band head not too strongly overestimated. Such a mismatch
has not been reported for any other CO band emission from evolved massive stars
and requires some further investigation.

When we compare the first CO band head observed in 2004 with the one observed in
2010 (see Fig.\,\ref{fig:s73-variab}) we see two important facts: The
intensity of the emission is definitely lower in 2010, but the width of the
band head itself remains constant. The latter implies that the rotation velocity
of the CO gas did not change, while the former suggests a change in column
density and/or size of the emitting area.

To test this hypothesis, we fixed both the CO temperature and rotation velocity
to the best-fitting values listed in Table\,\ref{tab:CO-model}. Then we modeled
the three observed spectral pieces individually, by purely modifying the column
density. This procedure is possible because the column density needed to fit
the observations is high, and optical depth effects become visible. This means 
that the shape of the band heads is determined by the column
density. In a purely optically thin scenario, the intensity of the CO band
emission would simply scale with the column density and a proper determination
of the column density would not be possible.

\begin{figure*}[t]
\begin{center}
\includegraphics[width=\hsize,angle=0]{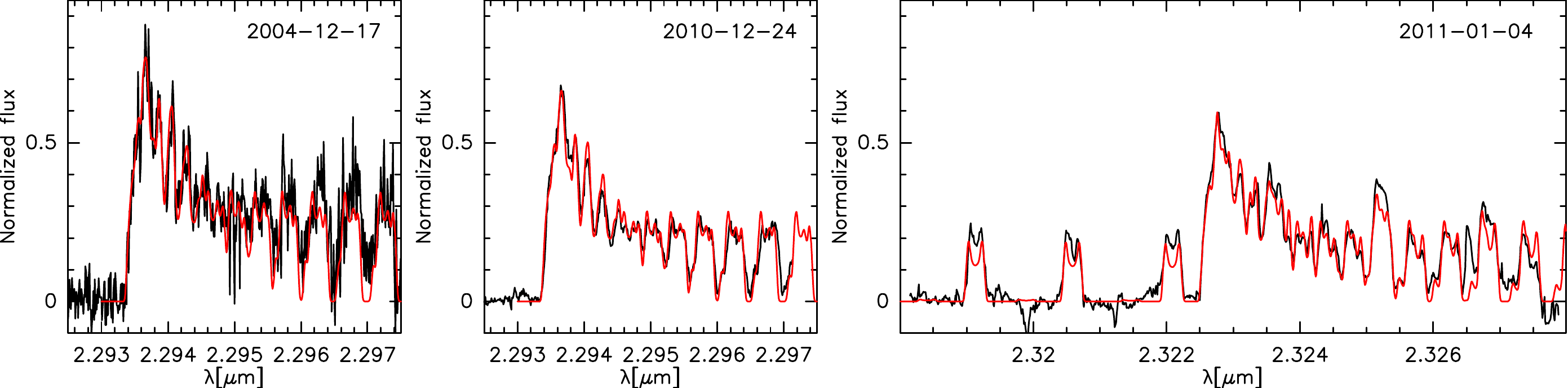}
\caption{Model fits (red) to the normalized and continuum-subtracted
(black) observations. Each model has a different column density and emitting area.}
\label{fig:S73-CO-comp}
\end{center}
\end{figure*}

The best-fitting parameters for the three individual spectral pieces are given 
in Table\,\ref{tab:CO-variab}, where we list the year of the observation (Col. 
1), the CO column density value (Col. 2), and the fitting parameter $f$ (Col. 
3, in arbitrary units). The latter parameter basically represents a normalized 
emission area. The larger errors in the parameters to fit the data from 2004 
are due to the lower quality of the spectrum. The resulting fits are shown in 
Fig.\,\ref{fig:S73-CO-comp}. The fits to the individual spectra from 2010 and 
2011 are considerably better than the one to the combined data set shown in 
Fig.\,\ref{fig:s73-CO-fit}. 
For comparison purpose, we list the values obtained from the fit to the 
combined observations (Fig.\,\ref{fig:s73-CO-fit}) in the last row of 
Table\,\ref{tab:CO-variab}.

%%%%%%%%%% Fig. 4 %%%%%%%%%%%%%%%%%%%%%%%%

%%%%%%%%%% Tab. 3 %%%%%%%%%%%%%%%%%%%%%%%%

\begin{table}
\caption{Density and fitting parameters.}
\label{tab:CO-variab}
\centering
\begin{tabular}{ccc}
\hline
\hline
Observation & $N_{\rm CO}$ & $f$   \\ \smallskip
            & ($10^{20}$\,cm$^{-2}$)  & (a.u.) \\ 
\hline

\smallskip

2004        & $7\pm 1$   & $3.65_{-0.11}^{+0.14}$ \\ \smallskip

2010        & $5\pm 0.5$ & $3.46_{-0.10}^{+0.13}$ \\ \smallskip

2011        & $4\pm 0.5$ & $3.01_{-0.12}^{+0.15}$ \\ \smallskip

2010$+$2011 & $6\pm 1$   & $2.94_{-0.14}^{+0.20}$ \\ 

\hline
\end{tabular}
\end{table}

%%%%%%%%%%%%%%%%%%%%%%%%%%%%%%%%%%%%%%%%%%%%%

%%%%%%%%%% Fig. 5 %%%%%%%%%%%%%%%%%%%%%%%%

\begin{figure}[t]
\begin{center}
\includegraphics[width=\hsize,angle=0]{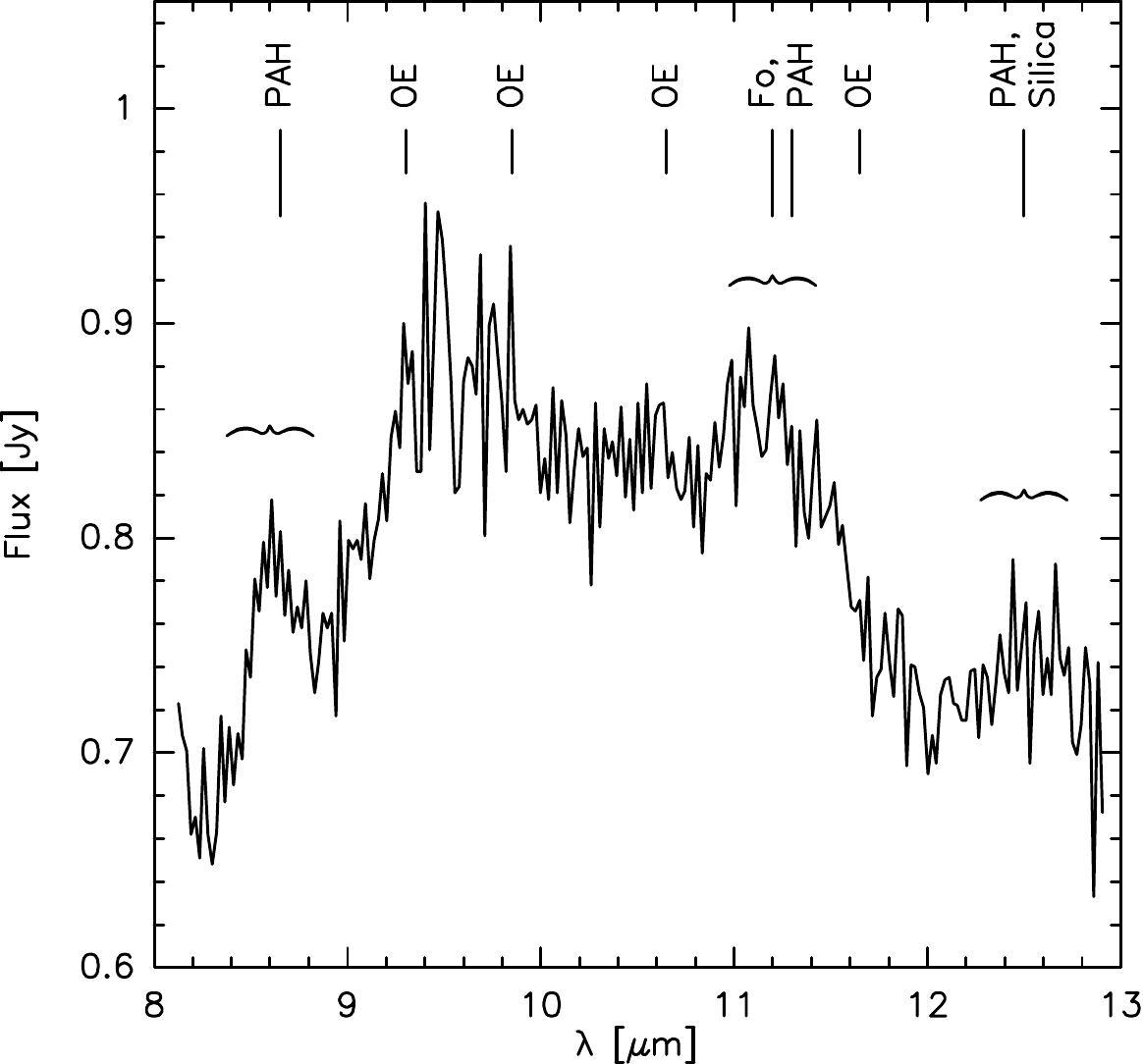}
\caption{Flux-calibrated mid-infrared spectrum. Prominent emission features
are identified: PAH -- polycyclic aromatic hydrocarbons, OE -- ortho enstatite,
Fo -- forsterite, and silica.}
\label{fig:s73-mir}
\end{center}
\end{figure}

\subsection{Dust emission}

The mid-infrared spectrum spans from 8\,$\mu$m to 13\,$\mu$m (see 
Fig.\,\ref{fig:s73-mir}). Its shape and intensity are comparable to the 
{\it Spitzer} spectrum \citepads{Kastner2010}. It clearly shows a prominent 
broad emission structure and, on top, several sharp emission features. 
According to the dust emission templates of \citet{2005A&A...437..189V}, we 
tentatively ascribe the broad structure to emission from large amorphous 
silicates (olivine, pyroxene, and silica), and we identify the sharp features
as emission from crystalline silicates (forsterite, ortho enstatite) 
together with weak emission from PAHs. LHA\,120-S\,73 hence displays a 
dual-dust chemistry, in which oxygen- and carbon-based grain particles coexist.

\subsection{Optical emission features}

\begin{figure*}[th]
\begin{center}
\includegraphics[width=\hsize,angle=0]{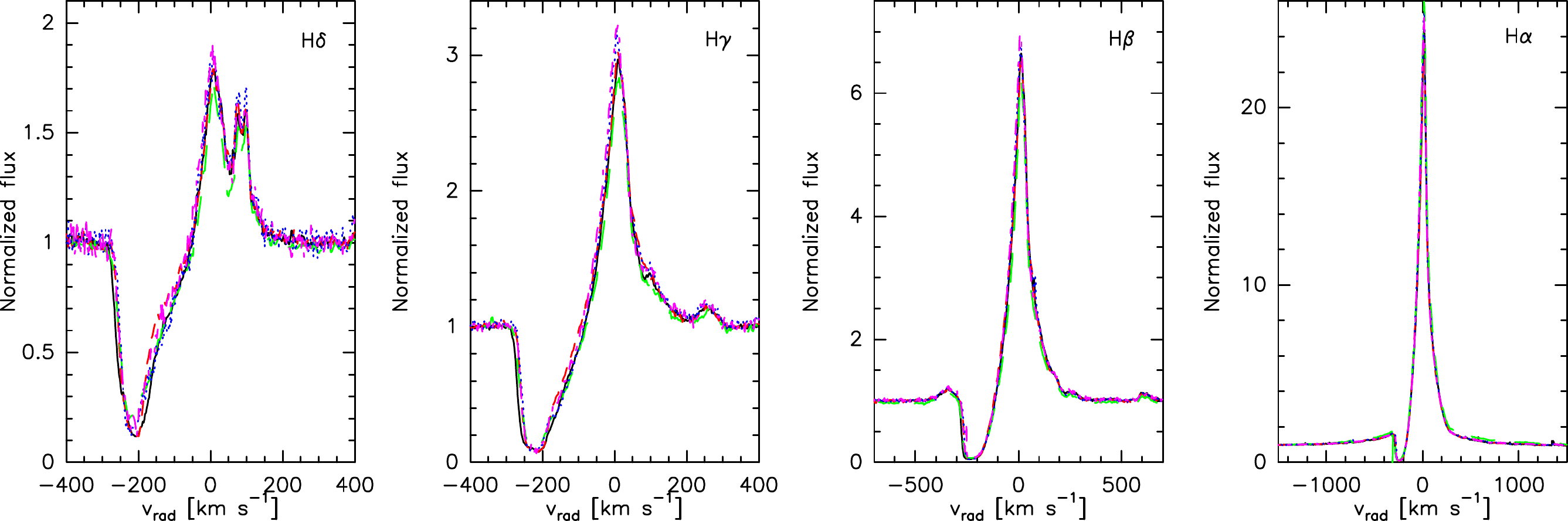}
\caption{Comparison of the strength and shape of the Balmer
lines of LHA\,120-S\,73 seen in FEROS spectra from 1999 (black, solid), 2005
(red, dashed), 2014 (blue, dotted), and 2015 (purple, dashed-dotted),
and in the du Pont data from 2008 (green, long-dashed).}
\label{fig:S73_Balmer}
\end{center}
\end{figure*}

\begin{figure*}[th]
\begin{center}
\includegraphics[width=0.75\hsize,angle=0]{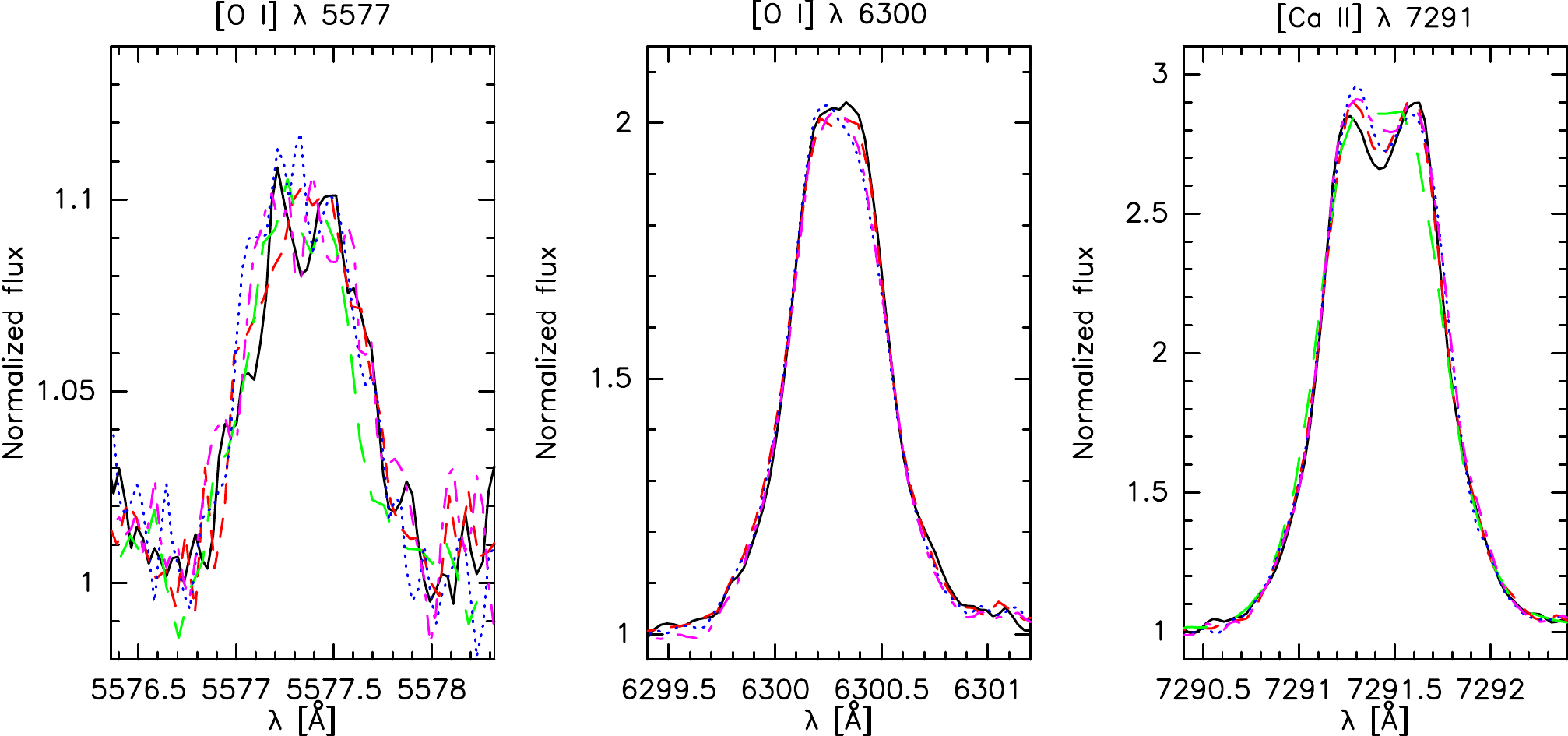}
\caption{Comparison of the strength and shape of some forbidden
lines of LHA\,120-S\,73 seen in FEROS spectra from 1999 (black, solid), 2005
(red, dashed), 2014 (blue, dotted), and 2015 (purple, dashed-dotted), and in
the du Pont data from 2008 (green, long-dashed), except for the [\ion{O}{i}]
$\lambda$ 6300 line, which could not be corrected for telluric pollution.}
\label{fig:S73_forb_variab}
\end{center}
\end{figure*}

The optical spectra of LHA\,120-S\,73 are crowded with countless narrow 
emission lines, particularly in the red portion. 
Most of these lines belong to permitted and forbidden transitions of 
\ion{Fe}{ii}. Only a few photospheric absorption lines are seen. We refrain 
from repeating the description of the optical spectra because they were 
described in great detail by \citetads{Stahl1983, 1985A&AS...61..237S}, 
\citetads{1986A&A...163..119Z}, and \citetads{1988A&A...190..103M}. 
Moreover, \citetads{1986A&A...163..119Z} reported that no indication for 
spectroscopic variability was seen in data taken sporadically over two decades.
The same statement is true for our data. The spectral features show 
no significant changes over the 16-year period covered by our observations. This
spectral stability can be seen in Figs.\,\ref{fig:S73_Balmer} and 
\,\ref{fig:S73_forb_variab}, where we compare the shape and strength of the 
Balmer lines and of different forbidden emission lines, respectively. As the 
CASLEO spectra have considerably lower resolution, we show here only the 
high-resolution (FEROS and du Pont) observations. In 
general, the line profiles from the different epochs agree very well. However, 
we observe small variations in the absorption component of the Balmer 
lines. In addition, the double-peaked [\ion{Ca}{ii}] $\lambda 7291$ line shows 
variability in the peak intensities (right panel of
Fig.\,\ref{fig:S73_forb_variab}): while the profile had a more intense red peak 
in 1999, it displays a more intense blue peak in 2014. In the other years
(2005 and 2015) the profile appears symmetric.

The [\ion{O}{i}] $\lambda 5577$ line also seems to display double-peaked 
profiles, although this line is very weak and hence very noisy, which distorts 
the real profile shape. The [\ion{O}{i}] $\lambda 6300$ line profiles appear 
single peaked at first glance, but we also note slight asymmetries.
The [\ion{O}{i}] $\lambda 6300$ line from the du Pont spectra was omitted from
Fig.\,\ref{fig:S73_forb_variab} because the shape and strength of its profile
is strongly affected by telluric pollution.

%%%%%%%%%% Fig. 6 %%%%%%%%%%%%%%%%%%%%%%%%

%%%%%%%%%% Fig. 7 %%%%%%%%%%%%%%%%%%%%%%%%

\subsubsection{Kinematics of the forbidden lines}

The forbidden emission lines of [\ion{Ca}{ii}] and [\ion{O}{i}] are of 
particular interest because they were suggested to originate in the high -density Keplerian disks of B[e]SGs \citepads{2007A&A...463..627K, 
2010A&A...517A..30K, 2012MNRAS.423..284A}. Moreover, these forbidden lines were 
found to trace different density regimes, with the [\ion{Ca}{ii}] lines forming 
closest to the star, the [\ion{O}{i}] $\lambda 5577$ line at similar or 
slightly larger distances, and the [\ion{O}{i}] $\lambda 6300$ line forming 
even farther out. In a recent study, \citetads{2016MNRAS.456.1424A} found that 
the profiles of the [\ion{Ca}{ii}] and [\ion{O}{i}] lines in the two 
investigated Galactic B[e]SGs (3\,Pup and MWC\,349) can be reproduced assuming 
that each emission originates from an individual Keplerian ring 
tracing a different distance from the star. When combining the information
from different atomic and molecular disk tracers for 3\,Pup, it became clear
that the disk around this star cannot be continuous but must be confined 
into seperate rings, at least one atomic and one molecular. This conclusion was 
drawn from the lack of hot CO band head emission \citep{Kraus2016}.

%%%%%%%%%% Fig. 8 %%%%%%%%%%%%%%%%%%%%%%%%

\begin{figure}[th]
\begin{center}
\includegraphics[width=\hsize,angle=0]{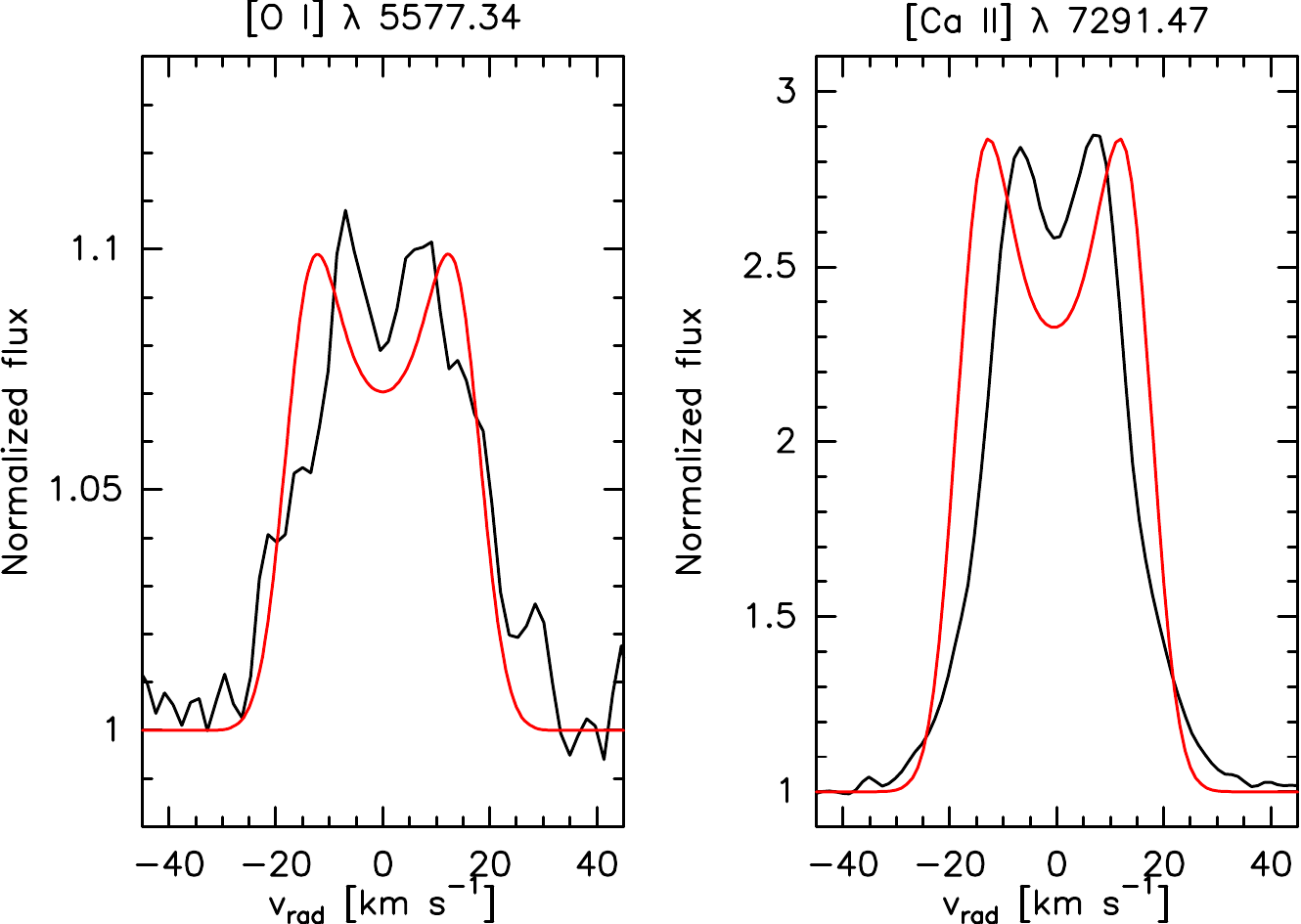}

\smallskip

\includegraphics[width=\hsize,angle=0]{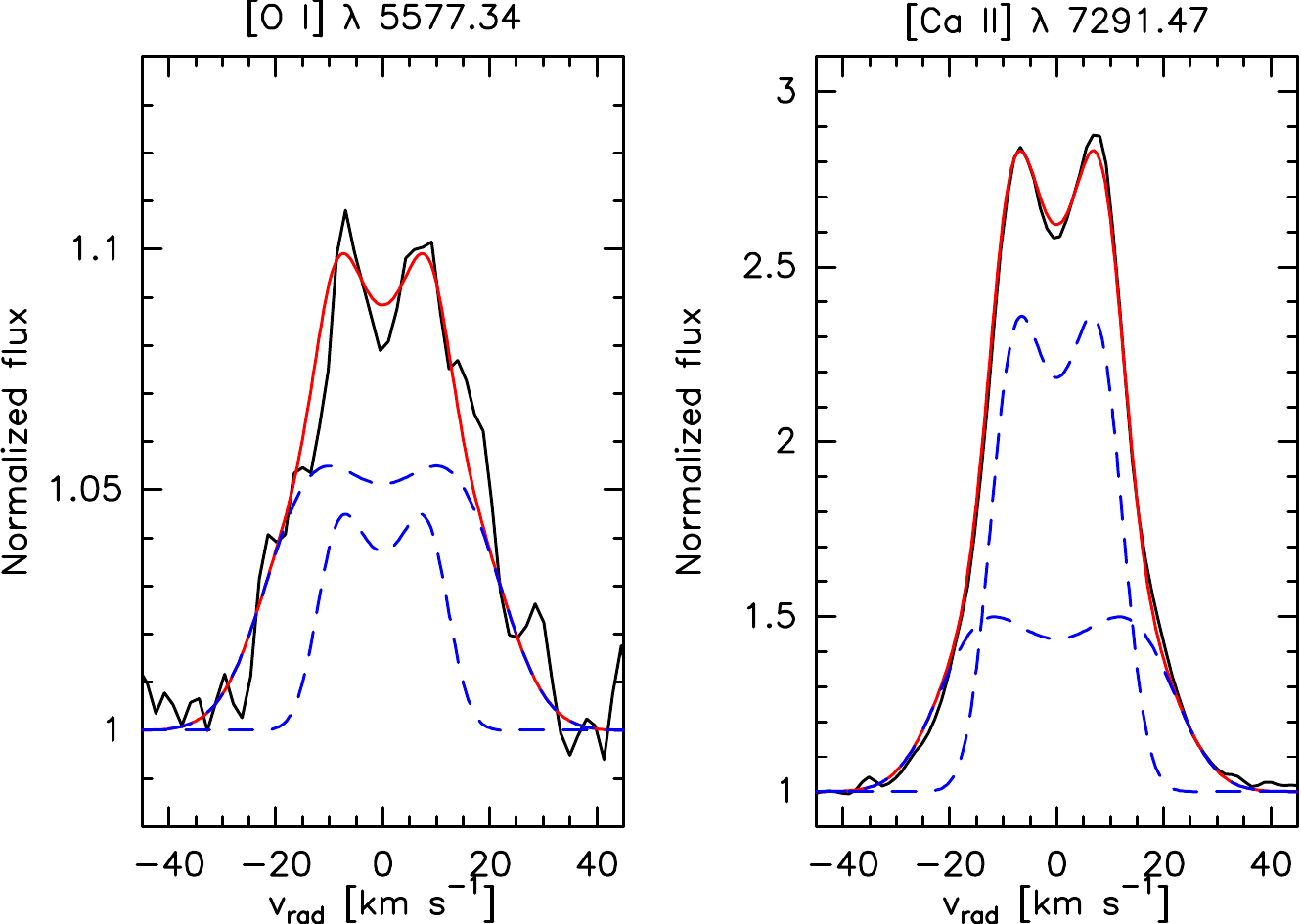}
\caption{Comparison of the observed line profiles seen in the FEROS 
spectrum from 1999 (black) with theoretical model fits (red) consisting of a 
single (top) and two Keplerian rings (bottom). In the latter, the blue lines 
show the contributions of the individual rings.}
\label{fig:S73_forb_ring}
\end{center}
\end{figure}

The low CO temperature of only 2800\,K obtained from our models suggests that 
the molecular gas disk around LHA\,120-S\,73 also has a sharp inner edge traced 
by the CO bands, indicative of an abrupt change in density. If the emission of 
the forbidden lines is confined to
a narrow gas ring, then either this ring should agree with the position of the
molecular gas, or it should be clearly detached from it. To study the kinematics
stored in the profiles of the forbidden emission lines in LHA\,120-S\,73, we
followed the approach of \citet{2016MNRAS.456.1424A} and applied a purely
kinematical model. We assumed that the emission is generated within a ring of
gas with a rotational and a Gaussian velocity component. The latter contains a
combination of at least the thermal velocity of $\sim 1-2$\,km\,s$^{-1}$ 
and the spectral resolution of FEROS of $\sim 5.5-6.5$\,km\,s$^{-1}$. An 
additional component due to some internal (turbulent) motion of the gas might 
be present as well, but will be neglected at first.

The [\ion{Ca}{ii}] $\lambda$ 7291 line and the [\ion{O}{i}] $\lambda$ 5577 line
display a clear double-peaked profile.
For a test computation, we used the CO emitting region as starting point and 
computed the profile resulting from a ring with a rotational velocity of 
34\,km\,s$^{-1}$ and a Gaussian component of 6.0\,km\,s$^{-1}$. This 
profile was then compared to the observed double-peaked line profiles (see the 
top panels of Fig.\,\ref{fig:S73_forb_ring}). Obviously, a single Keplerian 
ring with this rotational velocity cannot reproduce the observations, whose 
peaks are much narrower at the top of the lines and significantly wider in 
their wings. 
Consequently, the observed line profile must consist of at least two 
components, one with a higher and one with a lower rotation velocity than the 
CO gas. Therefore, we next tested the possibility of two Keplerian rings.
We found that a combination of two rings 
can fit the observed double-peaked lines very well. The fits are shown in the 
bottom panels of Fig.\,\ref{fig:S73_forb_ring} and the values needed for the 
individual velocity components 
are listed in Table\,\ref{tab:forb_fit}. Interestingly, the emission from the 
ring at larger distance from the star displays no additional broadening beyond
spectral resolution, while the emission from the inner ring requires a velocity
component of $\sim 8-9$\,km\,s$^{-1}$ in addition to the thermal and spectral 
resolution broadening. While parts of this velocity might be caused
by intrinsic (turbulent) motion of the gas, we interpret this component as
an indicator that the inner ring is not confined but has a certain radial 
extension.

%%%%%%%%%% Tab. 2 %%%%%%%%%%%%%%%%%%%%%%%%

\begin{table}
\caption{Profile kinematics of the forbidden lines.}
\label{tab:forb_fit}
\centering
\begin{tabular}{lcccc}
\hline
\hline
Line & $v_{\rm rot, 1}$ & $v_{\rm gauss, 1}$ & $v_{\rm rot, 2}$ & $v_{\rm gauss, 2}$  \\
     & (km\,s$^{-1}$)   & (km\,s$^{-1}$)   & (km\,s$^{-1}$)  & (km\,s$^{-1}$) \\
\hline
$[$\ion{Ca}{ii}] 7291  & $39\pm 1$ & $10\pm 1$ & $22\pm 1$ & $5.5\pm 1$  \\
$[$\ion{O}{i}] 5577    & $38\pm 2$ & $11\pm 2$ & $22\pm 2$ & $5\pm 2$  \\
$[$\ion{O}{i}] 6300    & $33\pm 1$ & $11\pm 1$ & $16\pm 1$ & $7\pm 1$  \\
\hline
\end{tabular}
\end{table}

%%%%%%%%%%%%%%%%%%%%%%%%%%%%%%%%%%%%%%%%%%%%%

%%%%%%%%%% Fig. 9 %%%%%%%%%%%%%%%%%%%%%%%%

\begin{figure}[th]
\begin{center}
\includegraphics[width=\hsize,angle=0]{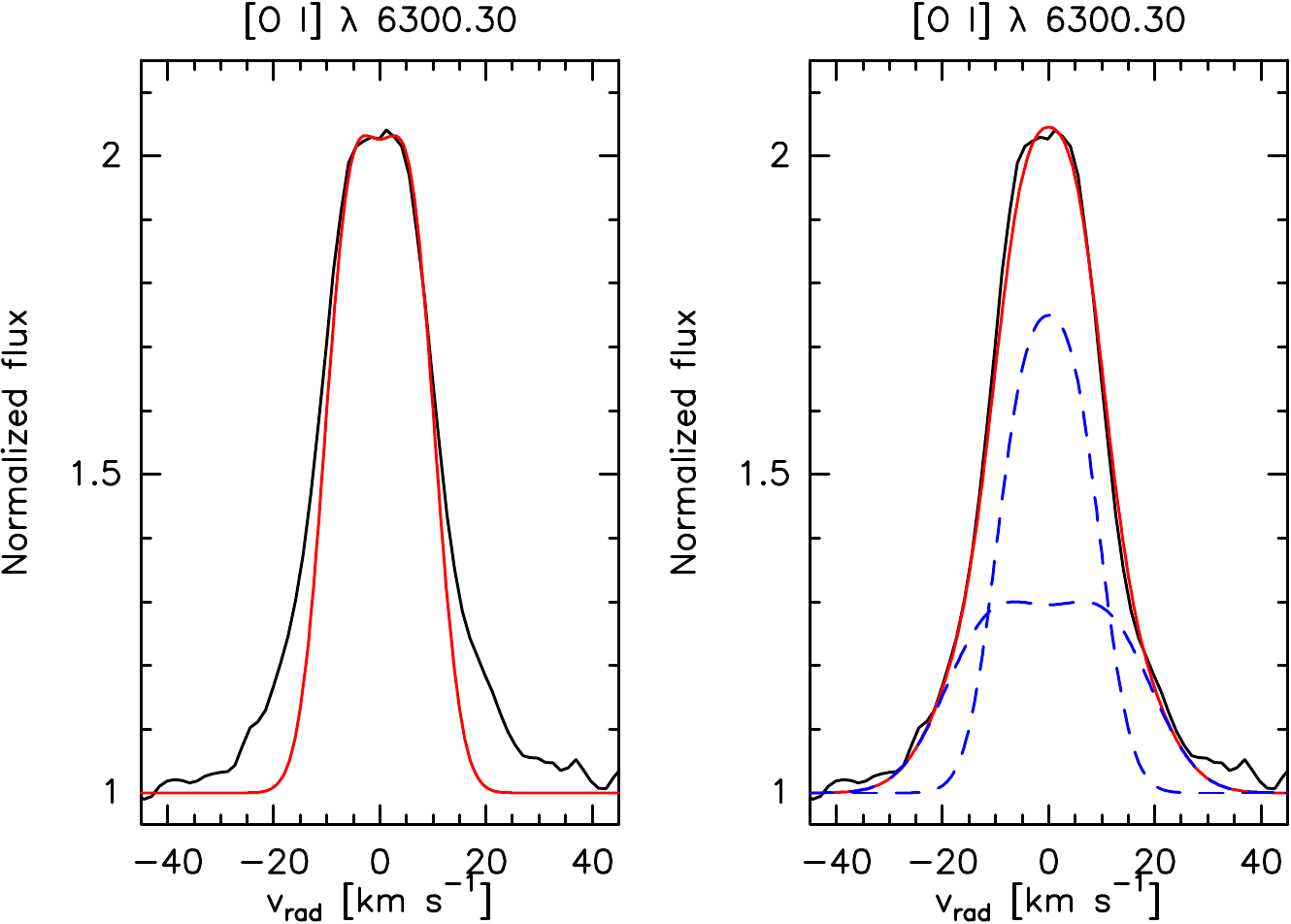}
\caption{Comparison of the observed line profile seen in the FEROS spectrum from 1999 (black) with theoretical model fits (red) consisting of a single 
(left) and two Keplerian rings (right). In the latter, the blue lines 
show the contributions of the individual rings.}
\label{fig:S73_forb_fit}
\end{center}
\end{figure}

%%%%%%%%%%%%%%%%%%%%%%%%%%%%%%%%%%%%%%%%%%%%%

Both double-peaked line profiles can be fit with basically identical 
velocity components, which agrees with earlier findings that these 
lines occur in regions of very similar density 
\citepads[e.g.,][]{2012MNRAS.423..284A}. The excitation conditions for the
levels from which the [\ion{Ca}{ii}] lines emerge are non-trivial. These
lines can be populated both through collisions from the ground level and through 
decay of energetically higher levels by the \ion{Ca}{ii} triplet line emission.
This makes it difficult to analyze the level population without a proper 
radiation transfer model for \ion{Ca}{ii}. The situation for the \ion{O}{i}
lines is easier. The levels from which the forbidden lines emerge are 
decoupled from the rest of the atom so that they are purely collisionally 
excited. As their emission is optimized within a rather narrow density range
\citepads[see, e.g,][]{2016arXiv160504066E}, the necessity of two rings 
suggests that the 
material in the environment of LHA\,120-S\,73 has radial density 
inhomogeneities. This means that at larger distance from the star similarly 
high densities must be achieved as closer to the star to have a sufficiently 
high level population of the levels from which the forbidden lines
emerge.

We now consider the [\ion{O}{i}] $\lambda$ 6300 line, which appears 
single-peaked in our spectra. This means that at least the central part of the 
line probably originates from a region with a lower Keplerian velocity component 
than the one of the second ring of the [\ion{Ca}{ii}] line, which still 
displays a double-peaked profile. A reasonable fit to this central
line profile region was obtained for a deprojected rotational velocity of 
$17.5\pm 0.5$\,km\,s$^{-1}$ and a Gaussian of 6.0\,km\,s$^{-1}$ (left panel of 
Fig.\,\ref{fig:S73_forb_fit}). However, as in the case of the double-peaked 
profiles, the broad wings cannot be fit with such a single ring. Therefore, 
we searched again for the best combination of two rings. The resulting fit is 
shown in the right panel of Fig.\,\ref{fig:S73_forb_fit} and the velocity 
parameters are included in Table\,\ref{tab:forb_fit}

Our results 
suggest that the emission in the [\ion{Ca}{ii}] and the [\ion{O}{i}] $\lambda 
5577$ lines originate from the same two rings with Keplerian velocities of 
38-39\,km\,s$^{-1}$ and 22\,km\,s$^{-1}$, respectively. In contrast, the 
first ring from which the [\ion{O}{i}] $\lambda 6300$ line originates is 
located in between these two [\ion{Ca}{ii}] rings and coincides with the 
location of the CO band head emitting ring, while the second is located 
much farther out (see also Sect.\,\ref{Sect:evolstate} and 
Table\,\ref{tab:ring-dist}).   

For the sake of completeness, we would like to add that we also considered 
the case that the forbidden emission lines might originate from a continuous 
disk. For this scenario assumptions need to be made for the radial density
distribution. A power law has proven to work 
reasonably well to reproduce the spectral energy distributions of 
pre-main-sequence stars' disks \citepads[see, e.g.,][]{1990AJ.....99..924B}. 
The density distribution within a stellar wind or an equatorial 
outflow can also be approximated by a power law \citep[with a different exponent, 
however, see][]{2007MNRAS.377.1343B, 2010A&A...517A..30K}. However, none of the
studied scenarios resulted in a reasonable fit to the observed line profiles
because they either were able to fit the broad wings but then produced too broad
profiles in the central line regions, or they were able to reproduce the central
regions but failed in producing broad enough line wings.

Considering that the double-peaked profiles have contributions from two
distinct regions (closer to the star and farther out than the CO ring)
that require similar densities, a continuous disk with a monotonic radial 
density distribution does not seem a very likely scenario for the
distribution of the circumstellar material of B[e]SGs. Instead, our model
results give preference to a scenario that involves multiple (sharp and/or
extended in radial direction) density concentrations, which could either 
indicate a strongly non-monotonic radial density distribution within a 
continuous disk or just individual rings of material. Support for such 
a scenario comes from recent imaging result of one Galactic B[e]SG, for which
multiple clumpy rings were resolved in the molecular CO gas 
\citepads{Kraus2016} and also in the atomic gas (Kraus et al., in preparation).

\subsubsection{TiO band emission}

\begin{figure}[t]
\begin{center}
\includegraphics[width=\hsize,angle=0]{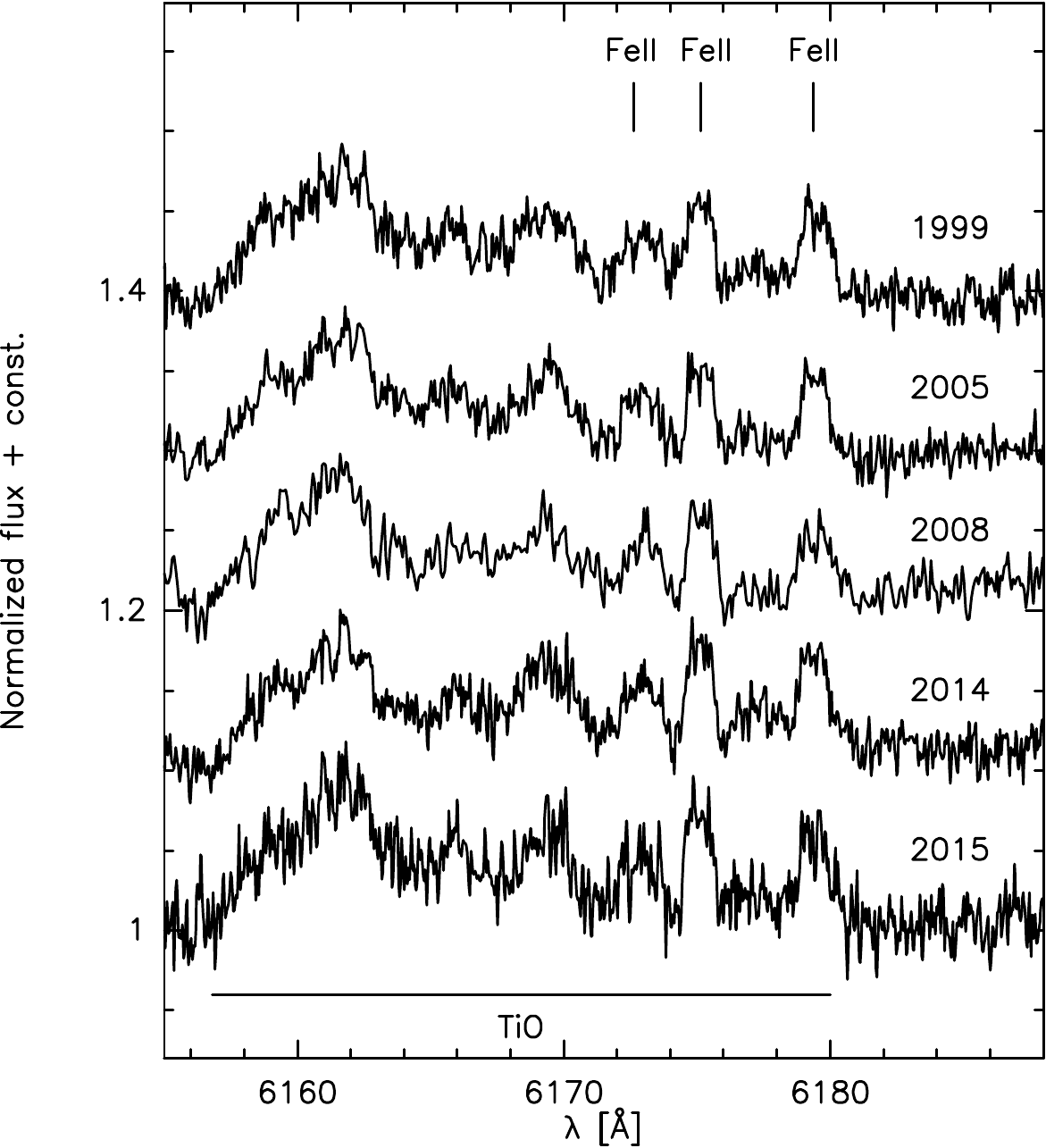}
\caption{Detection of TiO band emission in the high-resolution optical spectra of LHA\,120-S\,73.}
\label{fig:S73_TiO}
\end{center}
\end{figure}

The optical spectra display a broad emission feature extending from 
$\sim 6158$\,\AA \ to roughly 6180\,\AA \ (see Fig.\,\ref{fig:S73_TiO}). At the 
faint red end, some \ion{Fe}{ii} lines are superimposed, but the blue part of 
the feature is not affected by emission lines. Similar structures were 
clearly seen in three other B[e]SGs, the two LMC stars LHA\,120-S\,12 and 
LHA\,120-S\,134, and the Small Magellanic Cloud (SMC) object LHA\,115-S\,18, 
and were assigned to TiO band emission 
\citepads{1989A&A...220..206Z, 2012MNRAS.427L..80T}. 
LHA\,120-S\,73 is hence the fourth object with clear signatures of TiO band 
emission. Interestingly, all stars with TiO also display strong CO band 
emission \citepads[see, e.g.,][]{2013A&A...558A..17O}. This suggests that
the conditions in the environments (in terms of column density and emitting 
area) of these objects easily enable the production of measurable amounts of 
molecular emission. Hence emission from other species that form in oxygen-rich 
environments, such as SiO, which was recently discovered from four Galactic 
B[e]SGs \citepads{2015ApJ...800L..20K}, might also be detectable.

%%%%%%%%%% Fig. 9 %%%%%%%%%%%%%%%%%%%%%%%%

\section{Discussion}
\label{sec:disc}

\subsection{CO variability}

In Sect.\,\ref{sec:CO} we have shown that the observed variability in the 
CO band emission can result from simultaneous variations in the CO column 
density and the size of the emitting area. On the other hand, no noticeable 
change in rotation velocity of the gas ring was observed. 
Such a behavior cannot be explained with a 
dilution of the emission that is due to expansion of the ring, for example, because in this case 
the rotation velocity would be affected, that is, it would be
decreased. In addition, an
increase in emitting area together with a decrease in column density would be 
expected. However, the model fits do not agree with such a scenario. 
The fit to the spectrum taken in December 2010 requires a 25\% higher column 
density and simultaneously a 15\% larger emitting area than is
shown by the values 
obtained from fitting the data taken 11 days later (January 2011, see 
Table\,\ref{tab:CO-variab}). Therefore, we interpret the observed variability 
in the CO intensity as due to density inhomogeneities within the molecular ring.

%%%%%%%%%%% Fig. 10 %%%%%%%%%%%%%%%%%%%%%%%%
%
\begin{figure*}[t]
\begin{center}
\includegraphics[width=0.75\hsize,angle=0]{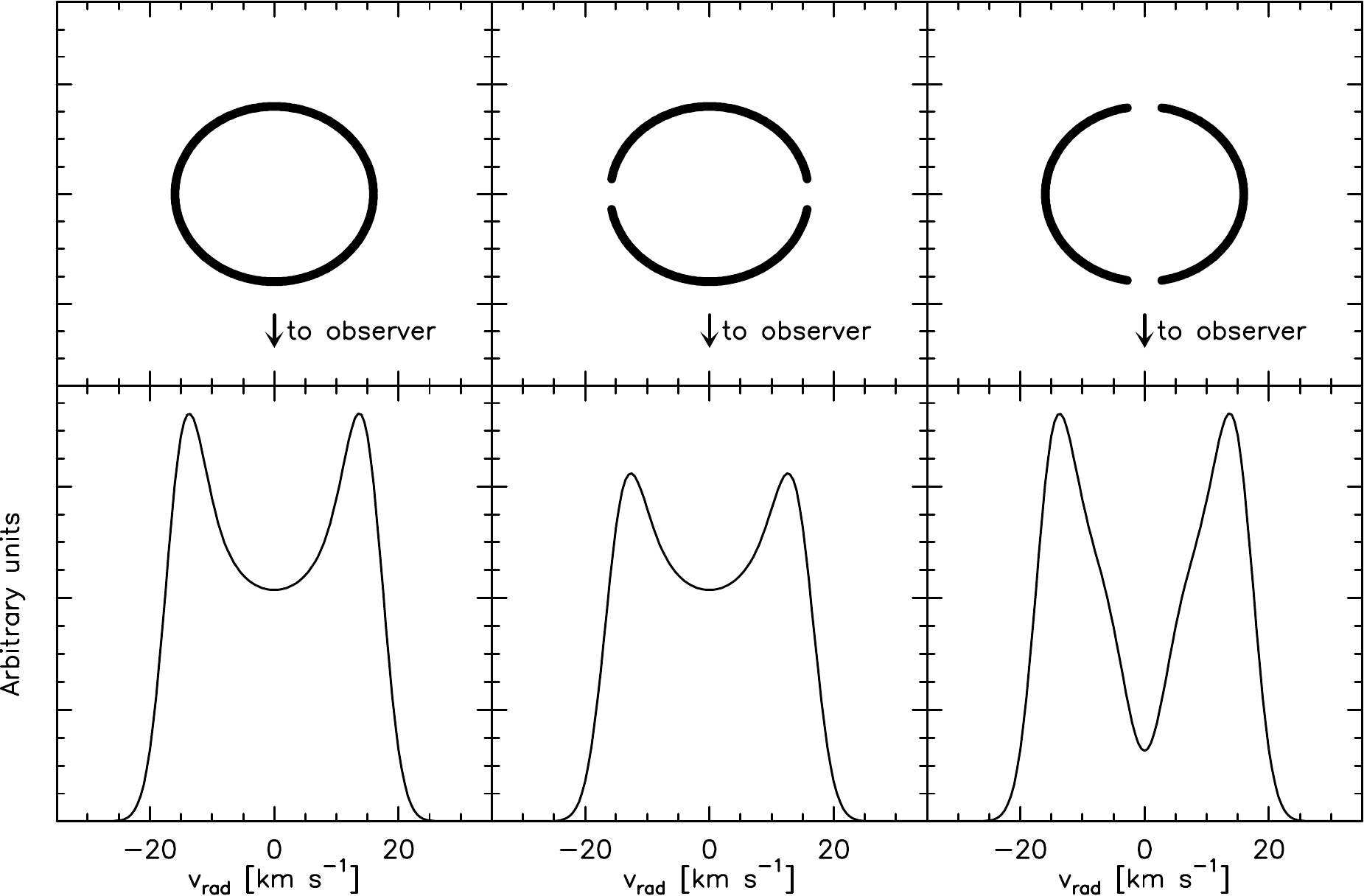}
\caption{Line profiles (bottom panels) resulting from different ring structures
(top panels): homogeneous (left) or with symmetric gaps (middle and right).}
\label{fig:s73-gaps}
\end{center}
\end{figure*}

We stress that the variability in CO intensity cannot result from the motion of 
the gas ring during 11 days, even if this ring has density inhomogeneities. 
Considering the mass estimate of 30\,M$_{\odot}$ \citepads{Stahl1983} for the 
initial mass of LHA\,120-S\,73 and the Keplerian rotation velocity of 
34\,km\,s$^{-1}$, the CO ring must have a radius of about 23\,AU and, hence, an 
orbital period of about 20\,years. Even if we assume that the current mass of 
LHA\,120-S\,73 is lower because the star experienced mass loss during its past
evolution (of $\sim 10-30\%$ of its initial mass), the orbital period would 
still exceed 10\,years. 
Therefore, 11 days are definitely too short for the motion of the gas to have a 
measurable influence on the line profiles.
The observed CO variability must hence result
from the intrinsic disposition of the inhomogeneities within the molecular 
ring.

To test this interpretation, we computed the line profiles of an optically 
thin gas ring with artificially implemented gaps. For simplicity and pure 
illustration purposes, we implemented two symmetric gaps into the ring, which 
otherwise remained unchanged (Keplerian rotation velocity of 34\,km\,s$^{-1}$,
inclination angle of 28\degr). The symmetry was suggested by the three resolved
rotation-vibrational lines in front of the second band head, which displayed no 
significant deviation from a symmetric double-peaked profile shape. Depending 
on the location and size of the gaps in the ring, the individual line profiles 
change substantially. This is demonstrated in Fig.\,\ref{fig:s73-gaps}, where 
we compare the profiles of a line formed in a homogeneous ring (left panel) 
with those formed in rings with symmetric gaps positioned at the 
highest (middle panel) and lowest (right panel) velocities 
and an angular size of 20$\degr$ each. The effect on the 
line profiles is clearly evident: while the peak intensities are suppressed when
the gaps are at the highest velocities, the central emission is 
strongly reduced in the model where the gaps are located at the lowest velocities. 

If we recall that the structure of the CO band heads results from the 
superposition of many individual rotation-vibrational lines \citepads[see, 
e.g.,][]{1995Ap&SS.224...25C}, it is obvious that the modifications and 
deformations of the individual profiles induced by the gaps considerably affect the shape and strength of the total CO band emission. 
Hence, to explain the deviations between the model fit and the observations,
the scenario of a clumpy structure of the molecular material seems to 
be most likely. We currently have no information on the 
number of clumps (or gaps), their sizes, surface densities, and positioning 
around the star at the time of observation. Such a clumpy ring 
structure has too many free parameters, hampering the proper modeling 
of the observed CO band spectra of LHA\,120-S\,73.

\begin{figure}[t]
\begin{center}
\includegraphics[width=\hsize,angle=0]{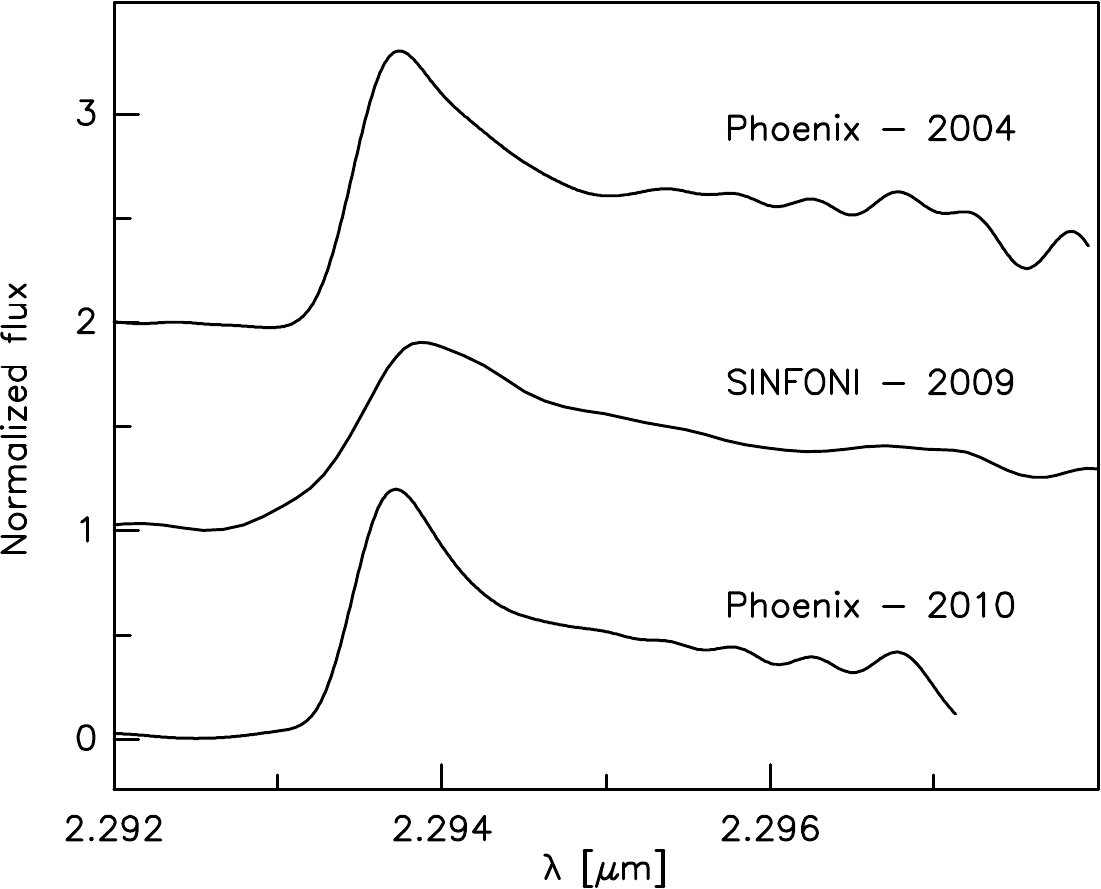}
\caption{Intensities of the first CO band head observed in different years.
For better comparison, the Phoenix spectra have been convolved to the
resolution of the SINFONI spectrum of \citetads{2013A&A...558A..17O}.}
\label{fig:s73-CO-conv}
\end{center}
\end{figure}

Further support that such a clumpy ring scenario could be reasonable is 
provided from a comparison of our results with those obtained recently 
by \citetads{2013A&A...558A..17O}. These authors observed LHA\,120-S\,73 in 
2009 as part of their $K$-band survey performed with the ESO medium-resolution 
($R\sim 4500$) integral-field unit spectrograph SINFONI. From their modeling 
of the CO band emission, they derived a CO temperature of $2800\pm 500$\,K that 
compares well to our value. However, the CO column density obtained by 
\citetads{2013A&A...558A..17O} is about a factor of six times higher than ours. 
We note that \citetads{2013A&A...558A..17O} modeled their spectrum without 
considering rotation of the CO gas because the structure of the CO band 
head did not display noticeable kinematical broadening beyond the spectral 
resolution of SINFONI. While the difference in the treatment of the underlying 
kinematics certainly influences the shape and strength of the observable CO bands 
(as we discussed above), the deviation is too large to have a purely kinematic origin. 
Therefore, we suspect that the intensity of the SINFONI CO band spectrum was 
different. To compare the intensities of the band head structures observed in 
different epochs, we convolved our high-resolution Phoenix spectra taken in 2004 and 2010 to the resolution of the SINFONI spectrum of 
\citetads{2013A&A...558A..17O} taken in 2009. The result is shown in 
Fig.\,\ref{fig:s73-CO-conv}. When we compare the intensities of the first 
CO band head emission, it is revealed that this has dropped from 2004 to 2009 and increased 
again in 2010 but remained below the intensity observed in 2004. This supports 
the idea that the molecular gas has a clumpy structure and revolves
around the star on stable 
Keplerian orbits.

%%%%%%%%%%% Fig. 11 %%%%%%%%%%%%%%%%%%%%%%%%

As was mentioned in Sect.\,\ref{sec:intro}, CO observations in B[e]SGs are 
rare, therefore not much is known about their possible variability. However, four
other objects were reported with clear changes in the molecular emission. The 
Galactic object CI\,Cam displayed strong CO band emission in observations taken 
one month after its spectacular outburst in 1998 \citepads{1999A&A...348..888C}.
This emission was no longer detectable in recent observations 
\citepads{2014MNRAS.443..947L}. The B[e]SG star LHA\,115-S\,65 in 
the SMC displayed a very stable and long-lived disk with no 
CO bands seen in the past decades, while \citetads{2012MNRAS.426L..56O} 
reported on the sudden appearance of strong CO band emission in that object. 
\citetads{2013A&A...558A..17O} detected CO band emission 
in the B[e]SG star LHA\,120-S\,35 in the LMC, 
while the observations of Torres et al. (in preparation) taken approximately 
two years later revealed that the intensity of the CO band emission in that 
object has decreased by about a factor of two.
The Galactic B[e]SG star HD\,327083 was reported to display CO band 
emission by \citetads{1988ApJ...324.1071M}. The CO bands were spatially 
resolved by \citetads{2012A&A...543A..77W}, who, based on interferometric 
observations combined with low-resolution $K$-band spectra, found no 
evidence for variability in CO within the observing period of five months. In 
contrast, high-resolution observations taken with a separation of only one 
month revealed noticeable differences in the shape of the CO band head 
\citepads{2013msao.confE.160K}. Moreover, fully resolved
rotation-vibrational lines in these data clearly display asymmetries in their 
double-peaked profiles: during the first observation the lines have a slightly 
stronger red peak, but they display a stronger blue peak one month later.
While the observed variability in CI\,Cam and LHA\,115-S\,65 {(and maybe also 
in LHA\,120-S\,35)} seems to be connected with highly dynamical phenomena 
(eruption, rapid expansion), HD\,327083 could be another example of a B[e]SG 
with a stable, clumpy molecular ring.

\subsection{Origin of the circumstellar rings}

Combining the information obtained from the atomic and molecular gas tracers,
it seems that LHA\,120-S\,73 is surrounded by multiple (at least four) gas rings. 
These rings populate distinct distances from the star, fixed by their Keplerian 
rotation speeds. Moreover, the fact that the first [\ion{O}{i}] $\lambda$ 6300 
ring is situated in between the two [\ion{Ca}{ii}] rings indicates that the 
density in the individual rings does not decrease from inside out. Instead, the 
two [\ion{Ca}{ii}] rings must have clearly higher densities than the 
[\ion{O}{i}] $\lambda$ 6300 rings, otherwise we would expect contributions of 
these ring regions to the [\ion{Ca}{ii}] line profiles as well.
In addition, the line profiles in our spectra taken over a period of 16 years
display no measurable change in width. This means that the emitting regions 
must be stable in position. The only noticeable variability in the line 
profiles of the forbidden emission lines occur in the peak intensities (see 
Fig.\,\ref{fig:S73_forb_variab}). This could indicate that the density in these 
rings also displays some inhomogeneities.

The formation mechanism of rings and disks revolving the B[e]SGs on stable 
Keplerian orbits is still a major unsolved problem \citepads[see][for a recent 
review]{2014AdAst2014E..10D}. The most commonly discussed scenarios are binary 
interaction up to full merger, and equatorially enhanced mass loss due to the 
rotationally induced bistability mechanism, which might be combined with a 
slow-wind solution. However, each of these scenarios has difficulties in 
reproducing all of the observed properties of B[e]SGs.

So far, indication for a companion was seen in only six B[e]SGs. Four of these 
stars (MWC\,300, HD\,327083, HD\,62623, and GG\,Car) are Galactic objects, and 
in all of them the disk appears to be circumbinary
 \citepads{2012A&A...545L..10W, 
2011A&A...526A.107M, 2012A&A...538A...6W, 2013A&A...549A..28K}, suggesting
that binary interaction might have caused the formation of these disks. The 
other two objects, LHA\,115-S\,6 and LHA\,115-S\,18, reside in the SMC. 
LHA\,115-S\,6 was suggested to be a post-merger object in an original triple 
system \citepads{1998ASSL..233..235L, 2006ASPC..355..259P}. This scenario was
proposed because it explains why the less massive component appears more 
evolved than the more massive one. The other object, LHA\,115-S\,18, was 
identified as the optical counterpart of a high-mass X-ray binary source
\citepads{2013A&A...560A..10C, 2014MNRAS.438.2005M}. Both SMC objects
display variabilities and features in their spectra that clearly separate them 
from the other B[e]SGs in the Magellanic Clouds, of which none displays 
indication for a companion. We cannot exclude that these remaining objects
are all merger remnants. However, in contrast to the strong variability 
and the diversity in spectral features observed in LHA\,115-S\,6, this scenario 
appears less likely.

The density enhancement in the equatorial plane achievable with the 
rotationally induced bistability mechanism is a factor 10--100 too low compared 
to the observations \citepads{2000A&A...359..695P}. When combined with the
slow-wind solution, the bistability mechanism is able to produce equatorial 
density enhancements of the right order \citepads{2005A&A...437..929C}. 
However, in that case, the wind velocities in the equatorial plane are vastly 
too high compared to the observed values. An additional hindrance 
with respect to LHA\,120-S\,73 is the relatively low stellar effective 
temperature of only 12\,000\,K, while the bistability jump occurs at 
temperatures around 22\,000\,K. A second bistability jump was suggested to 
occur around 10\,000\,K \citepads[see][]{2016MNRAS.tmp..168P}, suitable to 
enhance the mass-loss rate in metal-rich objects close to the Eddington limit. 
However, as a member of the LMC and with an initial mass of only $\sim 
30$\,M$_{\sun}$, LHA\,120-S\,73 does not necessarily fit into this group. 
Nevertheless, \citetads{2011ApJ...737...18C} found that in this low-temperature 
regime a slower and denser wind exists that fits the observed terminal 
velocities and the wind-momentum luminosity relation of A-type supergiants. 
Whether this new so-called delta-slow wind solution might also be relevant in 
the slightly hotter late B[e] supergiant stars still needs to be tested.

%%%%%%%%%%% Fig. 12 %%%%%%%%%%%%%%%%%%%%%%%%

\begin{figure}[t]
\begin{center}
\includegraphics[width=0.8\hsize,angle=0]{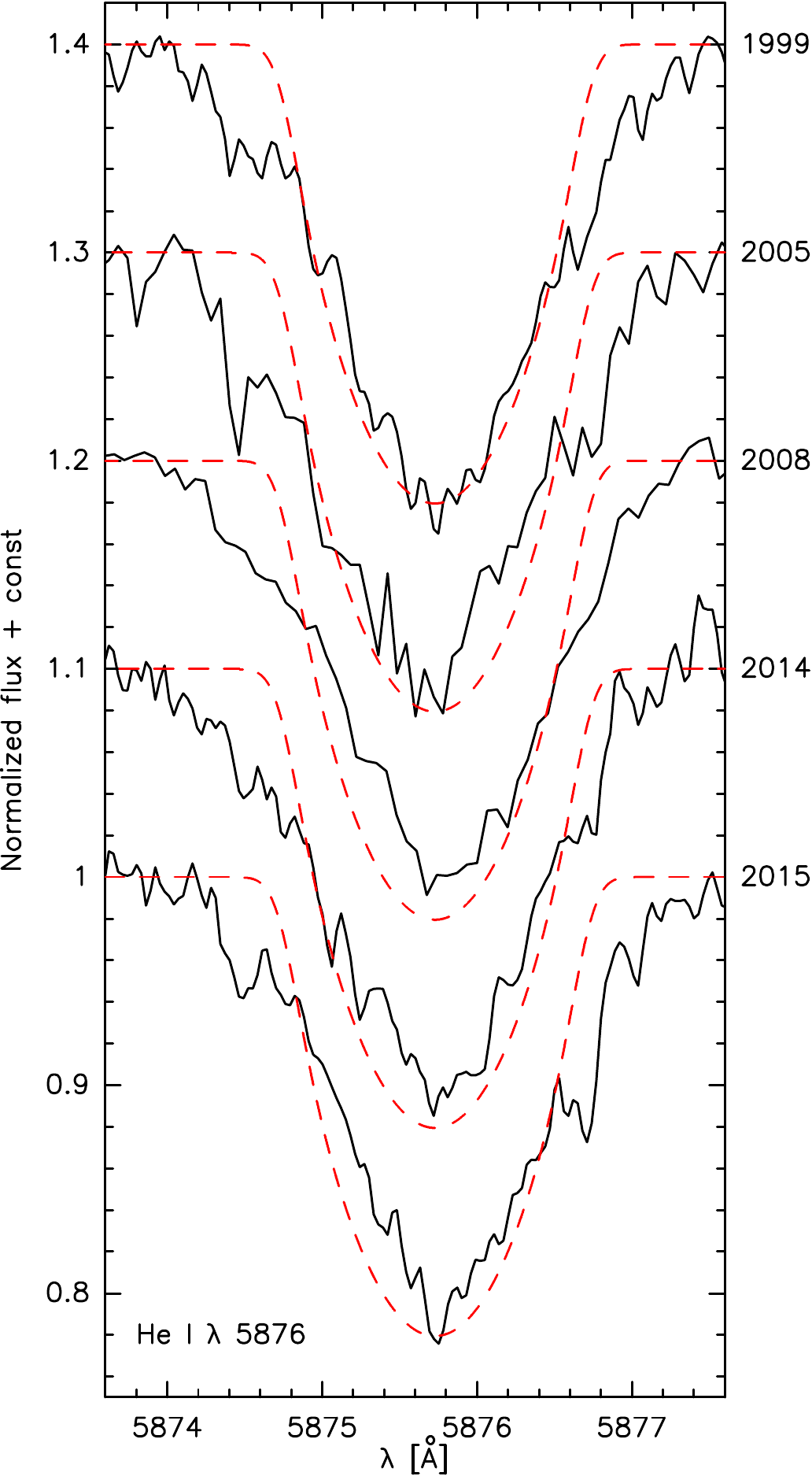}
\caption{Variability in the \ion{He}{i}\,$\lambda$5876 line.}
\label{fig:s73-HeI}
\end{center}
\end{figure}

Moreover, a prerequisite for the bistability mechanism is that the star should 
rotate at a substantial fraction of its critical velocity. LHA 120-S 73 is one 
of only four B[e]SGs for which the rotation velocity projected to the 
line of sight could be derived. \citetads{2006ASPC..355..135Z} estimated a 
value of $\varv\,\sin\,i \simeq 50$\,km\,s$^{-1}$ and a critical velocity of 
$\sim 140$\,km\,s$^{-1}$. Considering the inclination angle of 28$\degr$, the 
deprojected rotation velocity would be $\sim 106$\,km\,s$^{-1}$ and hence $\sim 75\%$ 
of the critical velocity. This value is similar to what was previously found 
for two B[e]SGs in the SMC, LHA\,115-S\,65 \citepads{2000ASPC..214...26Z, 
2010A&A...517A..30K} and LHA\,115-S\,23 \citepads{2008A&A...487..697K}, and 
would be sufficiently high for the bistability mechanism. However, we question
the accuracy of the value of $\varv\,\sin\,i \simeq 50$\,km\,s$^{-1}$.  

For the measurement of $\varv\,\sin\,i$ \citetads{2006ASPC..355..135Z} used 
the \ion{He}{i}\,$\lambda$5876 line, as it is the only obvious, pure 
atmospheric line seen in the spectrum. A few other lines are present as well, 
but they are too shallow and hence very noisy. 
We computed an artificial line profile,
considering only rotational broadening with $\varv\,\sin\,i =
50$\,km\,s$^{-1}$, thermal broadening for an effective temperature of
12\,000\,K, and broadening by the convolution with the spectral resolution. The
result is shown in Fig.\,\ref{fig:s73-HeI}, where we overplot the artificial
profile on those of the \ion{He}{i}\,$\lambda$5876 line observed in the
different epochs. Two obvious conclusions can be drawn from this comparison: 
(i) the observed line profiles look not as roundish as expected if they 
would originate from predominantly stellar rotational broadening, in contrast 
to the theoretical line profiles, and 
(ii) the \ion{He}{i}\,$\lambda$5876 line is variable in width and strength and
quite asymmetric.

This latter property is of particular interest. While we cannot exclude that 
the absorption line is polluted with emission from the wind and the 
circumstellar environment, the asymmetries strongly recall features visible 
in the profiles of non-radially pulsating stars \citepads[see, 
e.g.,][]{1997A&A...317..723T, 2010aste.book.....A}. Pulsations in 
LHA\,120-S\,73 were previously suggested by \citetads{2002A&A...386..926V}, who 
found indication for two periodic variabilities of $\sim 380$\,d and $\sim 
55.5$\,d in the photometric light curve, and by \citetads{2010AJ....140...14S},
who observed a period of 224\,days in the ASAS $V$-band light curve. Stellar 
pulsations can strongly 
influence the width and shape of atmospheric line profiles. Consequently,
the width of a line provides no reliable indicator for the contribution of 
stellar rotation, and we suspect that $\varv\,\sin\,i$ of LHA\,120-S\,73 is 
(much) smaller than 50\,km\,s$^{-1}$. If present, pulsations might also 
(quasi-)periodically enhance the mass loss of 
the star \citepads[see, e.g.,][]{2010A&A...513L..11A, 2015A&A...581A..75K}, 
so that during specific times density peaks might form in the radial outflow.
Depending on the pulsation modes, this "pulsed" mass loss might be confined 
within the equatorial plane, leading to ring structures in the outflow. 

On the other hand, these rings seem to be quasi-stationary and recall ripple 
structures because no noticeable expansion of the emitting material is seen in 
our spectra over the observing period of 16 years. Moreover, the sharpness of 
the observed line profiles, which indicates that the emission is confined to 
rather narrow rings, might imply that the space in between is mostly evacuated. 
This scenario strongly recalls ring systems around planets, in which moons 
have cleared the material in between the rings and stabilize the ring systems 
(so-called shepherd moons). Gaps populated by planets were also found in the 
accretion disks around pre-main-sequence objects \citepads[see, 
e.g.,][]{2015ApJ...815L..26R}. 

Although so far no planets were detected around evolved massive stars, 
the existence of one or more minor bodies revolving around LHA\,120-S\,73 and 
stabilizing the observed ring structures in analogy to the shepherd moons in 
planetary systems might offer a logical explanation. Observations of 
LHA\,120-S\,73 with the {\it Spitzer Space Telescope} IRS revealed an intense 
mid-infrared excess emission with clear indications of amorphous silicate dust. 
Moreover, specific spectral features originating from PAHs as well as from 
crystalline silicates were identified \citepads{2006ApJ...638L..29K} and are 
also seen in our mid-infrared spectrum (see Fig.\,\ref{fig:s73-mir}) taken 
about six years later. The 
coexistence of crystalline silicates and PAHs requires that grain processing in 
non-equilibrium chemistry has taken place in a long-lived, stable dust disk. 
It remains unclear whether this dust disk might have provided 
the material for the formation of possible planets (or minor bodies) or whether 
such objects might have formed simultaneously with the star and survived the 
previous evolution.

\subsection{Evolutionary state of LHA\,120-S\,73}
\label{Sect:evolstate}

The evolutionary state of B[e]SGs has long been unclear. The presence of
large amounts of ejected material was often considered as a strong indication 
for a post-red supergiant (post-RSG) stage. However, the investigations of
\citetads{2013A&A...558A..17O} revealed that the chemical composition
of the ejecta, in particular the enrichment in $^{13}$C, better agrees with
a pre-RSG scenario. This is also the case for LHA\,120-S\,73, although it
shows the strongest enrichment in $^{13}$C.

%%%%%%%%%% Tab. 5 %%%%%%%%%%%%%%%%%%%%%%%%

\begin{table}[t!]
\caption{Ring distances for $M_{*} = 27\,$M$_{\sun}$.}
\label{tab:ring-dist}
\centering
\begin{tabular}{llcc}
\hline
\hline
Ring No. & Element(s) & $\varv_{\rm rot}$ & $r$ \\
         &            & (km\,s$^{-1}$)    & (AU) \\
\hline
1  & [\ion{Ca}{ii}] ([\ion{O}{i}] $\lambda$ 5577) & 39 & 15.7 \\
2  & [\ion{O}{i}] $\lambda$ 6300, CO              & 34 & 20.7 \\
3  & [\ion{Ca}{ii}] ([\ion{O}{i}] $\lambda$ 5577) & 22 & 49.5 \\
4  & [\ion{O}{i}] $\lambda$ 6300                  & 16 & 93.6 \\
\hline
\end{tabular}
\end{table}

%%%%%%%%%%%%%%%%%%%%%%%%%%%%%%%%%%%%%%%%%%%%%

Additional support for a less evolved state is provided by the spectral energy
distribution of the dust, which is clearly distinct from those of (post-)RSGs
\citepads[see, e.g.,][]{2006AJ....132.1890B}. Furthermore, LHA\,120-S\,73 is a point 
source in {\it Spitzer} IRAC/MIPS images without any diffuse nebulosity
\citepads{Kastner2010}.

Moreover, so far, only two photometric
periodicities were found that might be interpreted as pulsational activity.
Computations of \citetads{2013MNRAS.433.1246S} for stars with an initial mass 
of $\sim 25$\,M$_{\sun}$ have shown that only few pulsations with periods of 
weeks to months are expected in the effective temperature range of 
LHA\,120-S\,73, as long as the star is in the pre-RSG evolution.
If LHA\,120-S\,73 were a post-RSG, it would undergo many more
pulsations \citepads{2013MNRAS.433.1246S, 2014MNRAS.439L...6G}. 
It is unclear as yet whether LHA\,120-S\,73 is indeed pulsating in just
two modes. Nevertheless, based on the listed arguments, we may conclude 
that the star is most likely in a pre-RSG evolutionary state. 
Considering an initial mass of the object of 30\,M$_{\sun}$ and evolutionary
tracks for rotating stars at LMC metallicity \citepads{2005A&A...429..581M},
the current mass of LHA\,120-S\,73 in a pre-RSG state would be about
27\,M$_{\sun}$. With this mass, we can compute the distances of the individual
gas rings. The results are listed in Table\,\ref{tab:ring-dist}.

\section{Conclusions}
\label{sec:concl}

We investigated the structure and kinematics of the circumstellar environment 
of the B[e]SG star LHA\,120-S\,73 in the LMC based on combined sets of 
high-resolution optical and near-infrared spectroscopic data collected over a 
time interval of 16 and 7 years, respectively. 

The near-infrared spectra cover the region of the first and second CO band 
heads. The high spectral resolution revealed rotational broadening of the 
individual rotation-vibrational CO lines with a deprojected rotation 
velocity of 34\,km\,s$^{-1}$. In addition, we found that the CO band head 
emission displays intensity variations, which we interpreted as
density inhomogeneities within the molecular ring.

In the optical spectra we discovered an emission feature that we identified as
molecular emission from TiO. LHA\,120-S\,73 is hence the fourth B[e]SG with 
confirmed TiO band emission. In general, the optical emission features show
no or only little indication for variability, suggesting that both the wind and
the circumstellar material are in quasi-stationary state. The spectra 
display emission of the forbidden lines of [\ion{Ca}{ii}] and 
[\ion{O}{i}], which are diagnostic tracers for circumstellar rings or disks.
The lines have double-peaked profiles, which we interpreted, in analogy to
previous studies, as due to Keplerian rotation. Modeling of the profile shapes 
revealed that each line profile consists of emission from two individual, 
spatially clearly distinct rings. 

The optical spectra also reveal one clear absorption line, identified as 
\ion{He}{i} $\lambda$ 5876. This line was formerly used to 
determine $\varv\sin i$ of the star and to demonstrate that the star is rapidly
rotating. However, from our data we could not confirm the previously reported 
high value of $\varv\sin i$. Instead, our spectra show that this line is 
clearly variable and highly asymmetric, reminding of line profiles of 
non-radially pulsating stars. Although the $S/N$ of our data is too poor to 
favor a pulsational origin of the line asymmetries over line pollution due to 
emission, the presence of at least two periodic variabilities determined 
from the photometric light curve of LHA\,120-S\,73 provides evidence that 
pulsations might indeed play a non-negligible role in the dynamics of the 
stellar atmosphere.
 
In summary, we found that the global structure of the circumstellar material, 
traced by the forbidden emission lines and the CO molecular bands, consists 
of at least four distinct rings of alternating densities. Although we have currently no 
observational proof, we might speculate whether such a ripple structure of the 
circumstellar material might be caused by the presence of minor bodies (e.g. 
in the form of planets) that evacuated the space in between the rings, caused
density inhomogeneities, and stabilize the currently observable ring 
structures, in analogy to the shepherd moons within the ring systems of 
planets or the rings observed around pre-main sequence objects. Follow-up observations with reasonable time resolution are clearly 
needed to investigate in more detail the variability, in particular of the 
molecular emission, to obtain better constraints on the density structure of 
the environment of LHA\,120-S\,73.

\begin{acknowledgements}
We thank the anonymous referee for comments that helped to improve the draft.
This research made use of the NASA Astrophysics Data System (ADS) and of the 
SIMBAD database, operated at CDS, Strasbourg, France, and of Astropy, a 
community-developed core Python package for Astronomy 
\citepads{2013A&A...558A..33A}. 
M.K., G.M., and D.H.N. acknowledge financial support from GA\,\v{C}R (grant 
number 14-21373S). The Astronomical Institute Ond\v{r}ejov is supported by the 
project RVO:67985815. This work was also supported by the institutional 
research fundings IUT26-2, IUT40-1, and IUT40-2 of the Estonian Ministry of Education and 
Research. L.C. acknowledges financial support from the Agencia de Promoci\'on 
Cient\'{\i}fica  y Tecnol\'ogica (Pr\'estamo BID PICT 2011/0885), CONICET 
(PIP 0177), and the Universidad Nacional de La Plata (Programa de Incentivos 
G11/137), Argentina.
Financial support for International Cooperation of
the Czech Republic (M\v{S}MT, 7AMB14AR017) and Argentina (Mincyt-Meys ARC/13/12 and CONICET 14/003) is acknowledged.
M.C. acknowledges the support from FONDECYT project 1130173 and
Centro de Astrof\'isica de Valpara\'iso, and R.H.B. acknowledges financial support from FONDECYT Project 1140076.
The observations obtained in 2014 and 2015 with the MPG 2.2m telescope were 
supported by the Ministry of Education, Youth and Sports project - LG14013 
(Tycho Brahe: Supporting Ground-based Astronomical Observations). We would like
to thank the observers (S. Ehlerova, A. Kawka) for obtaining the data.
\end{acknowledgements}

\bibliographystyle{aa} % style aa.bst
\bibliography{s73} % your references Yourfile.bib

\end{document}